\documentclass[aps,prb,twocolumn,longbibliography,superscriptaddress,amsmath,amssymb,floatfix,showpacs]{revtex4-2}
\usepackage{amsmath}
\usepackage{amssymb}
\usepackage{bm}
\usepackage{color}
\usepackage[utf8]{inputenc}
\usepackage{pifont}
\usepackage[colorlinks,linkcolor={blue},citecolor={blue},urlcolor={blue}]{hyperref}
\usepackage{mathtools}
\usepackage{booktabs}
\usepackage{multirow}
\usepackage{physics}
\usepackage{relsize}
\usepackage{mathrsfs}
\usepackage{mathdots}

\begin{document}

\title{Vortex-Beam-Driven Dirac Materials: Impurity and Polarization Effects on Light-Induced Vortex and Edge States}

\author{Trevor W. Walsh}
\thanks{These authors contributed equally to this work.}
\affiliation{Department of Physics and Astronomy and Nanoscale and Quantum Phenomena Institute, Ohio University, Athens, Ohio 45701, USA}
\author{Eric E. Caldwell}
\thanks{These authors contributed equally to this work.}
\affiliation{Department of Physics and Astronomy and Nanoscale and Quantum Phenomena Institute, Ohio University, Athens, Ohio 45701, USA}
\author{Nancy P. Sandler}
\affiliation{Department of Physics and Astronomy and Nanoscale and Quantum Phenomena Institute, Ohio University, Athens, Ohio 45701, USA}
\author{Mahmoud M. Asmar}
\email{masmar@kennesaw.edu}
\affiliation{Department of Physics, Kennesaw State University, Marietta, Georgia 30060, USA}

\begin{abstract}
We study impurity scattering and polarization detuning in finite-size vortex-light-beam–driven massive Dirac systems. In finite geometries, circularly polarized vortex light opens a dynamical gap where topological edge states coexist with photoinduced multiply quantized vortex states. We analyze how finite-size effects, vorticity, and effective particle-hole symmetry manifest in the quasienergy spectrum, real-space states, and local density of states. We show that angular-momentum mixing due to localized impurities and impurity clusters reshape vortex states, while when produced by circular polarization, it leads to a gradual filling of the dynamical gap with bulk-derived states. Our results indicate that both vortex and edge signatures remain observable in the presence of impurities and realistic polarization deviations, providing guidance for experimental realizations.
\end{abstract}
\maketitle

\section{Introduction}
Periodic driving offers a dynamical and nonthermal knob for band structure, symmetry, and topology control in materials~\cite{flreview1,flreview2,flreview3,flreview4,flreview5}. Specifically, quasiparticles in irradiated matter hybridize with light leading to the emergence of Floquet bands and photon-dressed quasiparticle excitations absent in the parent equilibrium system~\cite{FloqTIReview}. Floquet engineering have inspired numerous theoretical predictions of light-driven phenomena, ranging from topology to transport and magnetism~\cite{FloquetTI,FloquetTI2,graphene-top-ins,Virtual-ph,mitraandoka1,ftrans1,ftrans3,martin3,Asmar2024}. Experimentally, Floquet bands have been observed via time-resolved and angle-resolved photoemission spectroscopy on the surface of topological insulators~\cite{FloqExp1,FloqExp2}, black phosphorous~\cite{BlackPhosphorus}, and graphene~\cite{TrARPSGraph}. Transport measurements have revealed signatures of long-lived Floquet states and the light-induced Hall effect in graphene~\cite{FloqTraspGraphen,FloqExp3}. Additionally, these states have been observed via light induced Stark shifts of exciton resonances in WS$_2$ and through the large modulation of the optical non-linearity in MnPS$_3$~\cite{Floq_WS2,Floq_WS22,Floquet_Modualation}.
  
Optical control of quantum matter has traditionally relied on intensity, frequency, and polarization; however, additional degrees of spatiotemporal tunability can arise from spatially structured beams. Vortex light beams (VLBs), in particular, carry orbital angular momentum (OAM) in addition to the spin angular momentum (SAM) associated with circular polarization (CP)~\cite{OVB-Allen,OVB-Review}. These beams are characterized by space dependent intensity profiles that vanish at the core due to their helical phasefront, thereby exhibiting a vortex structure~\cite{lightV3,OVL-Oneil}. 
Theoretical studies  of material systems subjected to structured light~\cite{OVBReview,babakVLB} have shown the possibility of VLBs use for magnetic skyrmions generation~\cite{OVBmagnet1}, OAM dependent current distributions in GaAs~\cite{OVBphotocurrents}, and second harmonic generation in two dimensional ($2$D) electron gasses~\cite{OVB2ndH}. Simultaneously, experiments have shown the use of the VLB's OAM as a spectroscopic probe of magnetism in matter~\cite{OVBmagnet2}, enhanced photovoltaic effect in MoS$_2$~\cite{lightVexp11}, OAM dependent photocurrents in WTe$_2$ photodetectors~\cite{lightVexp10} and current distributions in GaAs~\cite{lightVexp12,lightVexp9}.


%

Refs.~\cite{Lauren2025,MD1} establish that VLB-driven massive Dirac materials host multiply quantized electronic vortex states whose vorticity is fixed by the beam’s OAM. Furthermore, under CP the full nonperturbative Floquet Hamiltonian conserves total angular momentum, which can be used to classify the Floquet states~\cite{Lauren2025}. 
However the study of the ideal unterminated system does not consider that (i) the CP component of a VLB should generate topological edge modes that appear in terminated samples (and can coexist with in-gap vortices), (ii) deviations from CP may induce angular-momentum mixing, thus testing the robustness of the in-gap states, and (iii) impurities likewise mix angular momentum, potentially obscuring their experimental signatures.

In this paper, we address these three issues by extending the Floquet formalism of Ref.~\cite{Lauren2025} and develop a fermion-doubling–free, Floquet-based numerical scheme for finite-size Dirac materials under VLB irradiation, including their concomitant topological edge states. The approach enables standard perturbation theory in a Floquet basis classified by total angular momentum. Within this framework, we quantify angular-momentum mixing induced by detuning from CP or by the presence of impurities, and we quantify its impact on the quasienergy spectrum, eigenstates, and the local density of states (LDOS), explicitly linking these perturbative effects to the system’s symmetries. Hence, the ultimate goal of this work is to provide more realistic descriptions to guide the experimental observation of the phenomena enabled in VLB-driven Dirac systems. 
\begin{figure}[ht!]
  \centering
  \includegraphics[width=0.5\textwidth]{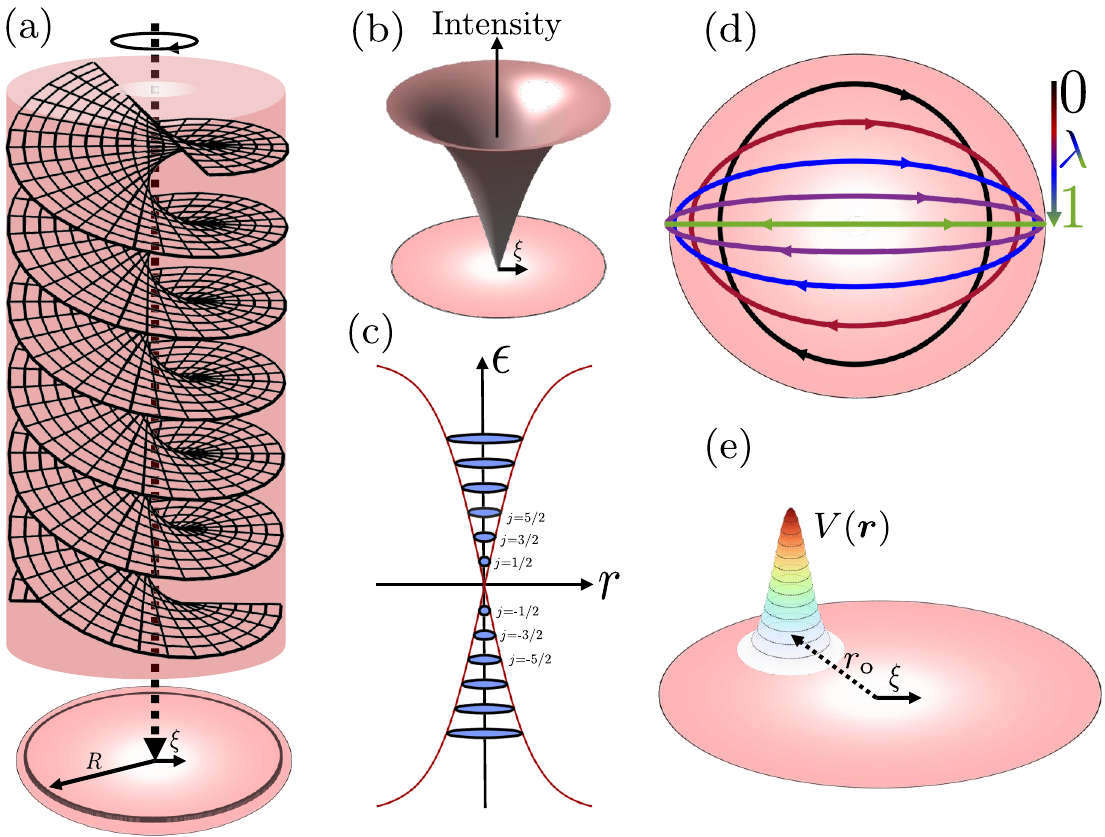}
  \caption{(a) Schematic of a VLB normally incident on a massive Dirac sample of radius $R$. The wavefront indicates the helical phase twist, and (b) shows the radial intensity profile of the VLB across the sample. (c) Bound vortex states as a function of energy and local intensity (characteristic radius of each state is indicated). (d) Polarization detuning: the drive's polarization is tuned from circular toward linear at fixed intensity. (e) Disorder: response in the presence of a Gaussian impurity located at radius $r_0$ from the sample center.}
  \label{Fig1}
\end{figure}

The remainder of this paper is organized as follows. Sec.~\ref{Sec1} presents the VLB-driven massive Dirac model and its Floquet symmetries, including total angular momentum and particle-hole symmetry. Sec.~\ref{Sec3} describes the fermion-doubling--free numerical method implemented to obtain the quasienergy spectra. This section also includes the analysis of the pristine spectrum, vorticity, edge and vortex states, and LDOS. Sec.~\ref{Sec4} examines impurity and polarization-detuning effects. Sec.~\ref{Disc} gives the conclusions and experimental outlook.

\section{Model and Formulation}\label{Sec1}
In this section we describe VLB driving of massive Dirac materials, present the model, and formulate its symmetries in the weak light-matter coupling regime.
\subsection{Modeled System} 
{\it Matter.}
We consider a $2$D massive Dirac material described at equilibrium by
\begin{equation}\label{H0}
H_{\rm D}(\bm r)= v_{\rm F}\,\boldsymbol{\sigma}\!\cdot\!{\bm p} +M\sigma_z,
\end{equation}
where $v_{\rm F}$ is the Fermi velocity, $2M$ is the semiconducting band gap, ${\bm \sigma}$ is the vector of the Pauli matrices acting on the spin degree of freedom, and $\bm p=-i\hbar{\bm \nabla}$ is the $2$D momentum operator. This material model can be realized on magnetically doped surfaces of topological insulators~\cite{MagTI1,MagTI2}, or thin films of these materials~\cite{thin2,thin4,thin3}.  

{\it Light.}
The material is irradiated by a VLB which is generally described by a vector potential of the form
\begin{equation}\label{Alight}
{\bm{\mathcal{{A}}}}({\bm r, t})=\Re\!\big\{A(r)\,e^{i(\Omega t-\ell\theta)}\,\boldsymbol{\varepsilon}_\eta\big\},
\end{equation}
where $\Omega$ is the drive frequency, $\theta$ is the azimuthal angle that determines the initial  phase of the vector potential and is measured in the same frame as the electronic degree of freedom [Fig.~\ref{Fig1}(a)]. ${\bm \varepsilon}_\eta=\hat{\bm x}+e^{-i\eta}\hat{\bm y}$ sets the polarization (tuning $\eta$ from $0$ to $\pi/2$ switches from linear to circular). For nonzero OAM  ($\ell\neq 0$) the azimuthal part of the field $e^{-i\ell\theta}$ imposes a $2\pi\ell$ phase winding, and hence a node along the propagation axis, $A(0)=0$, {\it i.e.}, a topological phase singularity, Fig.~\ref{Fig1}(a)--(b). For instance, in standard Laguerre–Gaussian (LG) beams with zero radial index, the intensity develops an annular maximum (“donut” profile)~\cite{OVB-Allen,OVB-Review,lightV3}. In the paraxial regime, the VLB envelope varies slowly compared to all lattice length scales. We assume a standard LG beam that saturates and remains bright within the sample, as it  decays well beyond the material's characteristic radius $R$. Hence, we model the radial profile of the VLB by
$A(r)=A_0\tanh(r/\xi)$, $A_0=E_0/\Omega$,
where $\xi<R$ sets the optical core scale over which the amplitude rises from zero to its saturation value $A_0$. This form effectively describes the beam as it averages the bright ring and captures both the near-axis null and radial saturation within the sample. For an LG beam with a wavelength $\lambda$ and waist $w_0\ge 2\lambda/\pi$, one expects a core diameter $2\xi\gtrsim \lambda/\pi$~\cite{lasers}. Hence, in this work we choose $\xi$ consistent with this bound and typical optical beam spot sizes that are within the range $1\sim 20 \;\mu\mathrm{m}$~\cite{lightVexp9,lightVexp11}. 

\subsection{General Formulation}
The electric field of the VLB couples to the equilibrium Dirac Hamiltonian [Eq.~\eqref{H0}] through minimal substitution,
${\bm p} \to {\bm p} + e {\bm{\mathcal{{A}}}}({\bm r, t})$, resulting in the space–time–dependent Hamiltonian
\begin{equation}\label{s-t-Hamiltonain}
\mathcal{H}(\mathbf r,t)
=
H_{\rm D}(\mathbf r)
+
e v_{\rm F}\,\boldsymbol{\mathcal A}(\mathbf r,t)\!\cdot\!\boldsymbol{\sigma}.
\end{equation}
Since $\mathcal{H}(\mathbf r,t)$ is periodic in time with period $T=2\pi/\Omega$, Floquet theory applies and yields solutions of the time-dependent Schrödinger equation for spinors
\begin{align}\label{states}
&
\hat{\Psi}_m(\mathbf r,t)
=
e^{-i\epsilon_m t/\hbar}\,\hat{\Phi}_m(\mathbf r,t),
\qquad
\epsilon_m \equiv \epsilon_m \ \mathrm{mod}\ \hbar\Omega,
\nonumber\\
&
\hat{\Phi}_m^{\mathrm T}(\mathbf r,t)
=
\big[\phi_{m,\uparrow}(\mathbf r,t),\phi_{m,\downarrow}(\mathbf r,t)\big],
\hat{\Phi}_m(\mathbf r,t+T)=\hat{\Phi}_m(\mathbf r,t),
\nonumber
\end{align}
where $m$ labels the Floquet sideband and $\epsilon_m$ denotes the quasienergy.
Substituting $\hat{\Psi}_m(\mathbf r,t)$ into the Schrödinger equation leads to
\begin{equation}\label{FLSC}
\big[\mathcal{H}(\mathbf r,t)-i\hbar\partial_t\big]\hat{\Phi}_m(\mathbf r,t)
=
\epsilon_m\,\hat{\Phi}_m(\mathbf r,t),
\end{equation}
where we define the Floquet operator $\mathcal{H}_{\rm F}(\mathbf r,t)\equiv\mathcal{H}(\mathbf r,t)-i\hbar\partial_t$.
Time-periodic Floquet modes may be represented in the extended Floquet–Sambe space~\cite{Floq-Sambe,Floq-Shirley} as
\begin{equation}\label{phiSambe}
\hat{\Phi}_m(\mathbf r,t)=P(t)\,\hat{\Phi}_m(\mathbf r),
\end{equation}
with a unitary, $T$-periodic “harmonic” operator
$
P(t)=P(t+T)=\exp(i\mathbb{N}_{\rm F}\Omega t),
$
and the Floquet number operator is
\begin{equation}\label{Nf}
\mathbb{N}_{\rm F}
=
\mathrm{diag}(..,n,n-1,..,1,0,-1,..,-n+1,-n,..).
\end{equation}
The time-independent Sambe vector is
\begin{equation}\label{Phi}
\hat{\Phi}^{\rm T}_m(\mathbf r)
=
[..,\hat{\phi}^{n}_m(\mathbf r),\hat{\phi}^{n-1}_m(\mathbf r),..,
\hat{\phi}^{-n+1}_m(\mathbf r),\hat{\phi}^{-n}_m(\mathbf r),..].
\end{equation}
Inserting Eq.~\eqref{phiSambe} into Eq.~\eqref{FLSC} and left–multiplying by $P^\dagger(t)$ yields
\begin{equation}\label{genHf}
\left[
P^{\dag}(t)\,\mathcal{H}(\mathbf r,t)\,P(t)
-
i\hbar\,P^{\dag}(t)\partial_t P(t)
\right]
\hat{\Phi}_m(\mathbf r)
=
\epsilon_m\,\hat{\Phi}_m(\mathbf r).
\end{equation}
Projecting onto the harmonic basis then gives the time–independent Floquet–Sambe matrix elements
\begin{eqnarray}\label{FHgeneral}
(\mathcal{H}_{\rm F})_{n'n}(\mathbf r)
=
\mathcal{H}_{n'n}(\mathbf r)
+
n\hbar\Omega\,\delta_{n'n},
\end{eqnarray}
where 
\begin{equation}\label{FMatrix}
\mathcal{H}_{n'n}(\mathbf r)
=
\frac{1}{T}\int_0^{T}
e^{i(n-n')\Omega t}\,
\mathcal{H}(\mathbf r,t)\,dt.
\end{equation}
are the Fourier components of the driven Hamiltonian.

For the VLB-driven system we use Eq.~\eqref{s-t-Hamiltonain} and obtain a block–tridiagonal structure in the harmonic index,
\begin{equation}\label{HFloquet}
(\mathcal{H}_{{\rm F}})_{n'n}(\mathbf r)= \big[H_{\rm D}(\mathbf r)+ n\hbar\Omega\big]\delta_{n',n}
+ \sum_{\gamma=\pm1}\mathcal{H}_{\gamma}(\mathbf r)\,\delta_{n',n+\gamma},
\end{equation}
where the diagonal blocks carry the static Dirac term and the photon ladder. The off–diagonal blocks describe single–photon absorption/emission and inherit the VLB’s OAM phase $e^{\mp i\ell\theta}$ and the generic polarization phase $\eta$:
\begin{equation}\label{Hpm}
\mathcal{H}_{\gamma}(\mathbf r) = \frac{\tilde{A}(r)}{2}\,e^{\bar{\zeta} i\ell\theta}
\Big[(1+ i \bar{\zeta} e^{i \bar{\zeta} \eta})\sigma_{\zeta}+(1+i \zeta  e^{i\bar{\zeta} \eta})\sigma_{\bar{\zeta}}\Big],
\end{equation}
$\zeta ={\rm sgn}(\gamma)$, with $\bar{\zeta}=-\zeta$, $\tilde{A}(r)=e\,v_{\rm F}A(r)$, and $\sigma_{\gamma}=(\sigma_x+i\gamma\sigma_y)/2$.
Equation~\eqref{HFloquet} is time independent but infinite dimensional in $n$, i.e., an infinite set of coupled, space-dependent differential equations for the components $\hat{\phi}_m^{\,n}(\mathbf r)$.

\subsection{Symmetries of the Floquet Hamiltonian}\label{FLsymmetry}
Here, we discuss the conservation of total angular momentum and the Floquet particle-hole symmetry, and how they constrain the spectrum and states.

Conservation of total angular momentum requires CP of the VLB $(\eta=\tau\pi/2,\ \tau=\pm 1)$~\cite{Lauren2025}. Under this condition the light–matter coupling is helicity-selective, such $\mathcal{H}_{\gamma}(\mathbf r)$ in Eq.~\eqref{Hpm} becomes 
\begin{equation}\label{cplight}
\mathcal{H}_{\gamma}(\mathbf r)
= \frac{\tilde{A}(r)}{2}\,e^{\bar{\zeta} i\ell\theta}
\Big[(1-\tau)\sigma_{\zeta}+(1+\tau)\sigma_{\bar{\zeta}}\Big].
\end{equation}
For circular driving there exists a Floquet–Sambe operator that commutes with $\mathcal{H}_{\rm F}$,
\begin{equation}\label{Jz2}
\widetilde{\mathcal{J}}_z^{\rm F}
=\left(\mathcal{L}_z\sigma_0+\frac{\sigma_z}{2}\right)\mathbb{I}_{\rm F}
+(\ell+\tau)\sigma_0\mathbb{N}_{\rm F},
\end{equation}
where $\tau$ sets the handedness of the VLB polarization,  
and $\mathbb{N}_{\rm F}$ is given in Eq.~\eqref{Nf}. The general Floquet Hamiltonian in Eq.~\eqref{FHgeneral} exhibits a Floquet particle-hole symmetry~\cite{PHFloquet} that, for our system, is defined by the antiunitary operator
\begin{equation}\label{PHop}
  \mathcal{P}=\sigma_{x}\mathbb{A}_{\rm F}K\;,
\end{equation}
where $K$ is the standard complex conjugation operator, and the unitary involution $\mathbb{A}_{\rm F}$ has matrix elements
\begin{equation}
\mathbb{A}_{sn} = e^{i\pi s}\,\delta_{s,-n},
\end{equation}
which reverses the Floquet harmonic index while imprinting a staggered $\pi$ phase, and satisfies $\mathbb{A}_{\rm F}\mathbb{N}_{\rm F}\mathbb{A}_{\rm F}=-\mathbb{N}_{\rm F}$. Using these relations, one verifies that $\{\mathcal{H}_{\rm F},\mathcal{P}\}=0$ for $\mathcal{H}_{\rm F}$ in Eq.~\eqref{HFloquet}. This implies that if $\mathcal{H}_{\rm F}\hat{\Phi}_{m}=\epsilon_{m}\hat{\Phi}_{m}$, then $\mathcal{H}_{\rm F}(\mathcal{P}\hat{\Phi}_{m})=-\epsilon_{m}(\mathcal{P}\hat{\Phi}_{m})$. Labeling the Floquet sidebands such that $\mathcal{P}\hat{\Phi}_{m}=e^{i\delta}\hat{\Phi}_{\bar m}$ ($\delta\in \mathbb{R}$) yields $\epsilon_{\bar m}=-\epsilon_{m}$, {\it i.e.} each state is paired with a partner at opposite quasienergy, independent of the VLB polarization. Importantly, when $\widetilde{\mathcal{J}}_z^{\rm F}$ is conserved, particle–hole symmetry imposes
$\widetilde{\mathcal{J}}_z^{\rm F}(\mathcal{P}\hat{\Phi}_{m,j})=-j(\mathcal{P}\hat{\Phi}_{m,j})$,
so that
$\mathcal{P}\hat{\Phi}_{m,j}=e^{i\delta}\hat{\Phi}_{\bar m,-j}$. However, the definition $\widetilde{\mathcal{J}}_z^{\rm F}$ is gauge dependent, and since we are interested in the phenomena near the Floquet-zone boundary it is convenient to work with, 
\begin{equation}\label{Jz2}
\mathcal{J}_z^{\rm F}
=\left(\mathcal{L}_z\sigma_0+\frac{\sigma_z}{2}\right)\mathbb{I}_{\rm F}
+(\ell+\tau)\sigma_0\left(\mathbb{N}_{\rm F}-\frac{\mathbb{I}_{\rm F}}{2}\right),
\end{equation}
where $\mathbb{I}_{\rm F}$ is the Floquet identity. These two definitions are related by a gauge transformation given by the unitary operator $\mathbb{U}=\exp[i\mathbb{I}_{\rm F}(\ell+\tau)\theta/2]\sigma_0$, such that $\mathcal{J}_z^{\rm F}=\mathbb{U}\widetilde{\mathcal{J}}_z^{\rm F}\mathbb{U}^{\dag}$. 

Working in this gauge, conservation of $\mathcal{J}_z^{\rm F}$ leads to the labeling of Floquet states by the eigenvalue $j$. Thus, we construct Floquet states in Eq.~\eqref{Phi} with total angular momentum $j$,
\begin{equation}\label{Phij}
\hat{\Phi}_{m,j}^\mathrm{T}(\mathbf r)
=\big[\ldots,(\hat{\phi}_{m}^{\,1})^\mathrm{T}(\mathbf r),(\hat{\phi}_{m}^{\,0})^\mathrm{T}(\mathbf r),(\hat{\phi}_{m}^{\,-1})^\mathrm{T}(\mathbf r),\ldots\big],  
\end{equation}
where the $n$th component of the Floquet spinor is
\begin{equation}
(\hat{\phi}_{m}^{\,n})(\mathbf r)
= e^{\,i[\ell_e-(\ell+\tau)n]\theta}\,\hat{\varphi}_{m}^{\,n}(\bm r),
\end{equation}
and $\hat{\varphi}_{m}^{\,n}(\bm r)$ written in its spin components is 
\begin{equation}
(\hat{\varphi}_{m}^{\,n})^\mathrm{T}(\mathbf r)
=\big[\phi^{\,n}_{m,\uparrow}(r)\,e^{-i\theta},\ \phi^{\,n}_{m,\downarrow}(r)\big]\;.
\end{equation}
Consequently, 
\begin{equation}\label{JphiValue}
\mathcal{J}_z^{\rm F}\,\hat{\Phi}_{m,j}(\mathbf r)
=\Big[\ell_e-\tfrac{1}{2}(\ell+\tau+1)\Big]\hat{\Phi}_{m,j}(\mathbf r)
= j\,\hat{\Phi}_{m,j}(\mathbf r),
\end{equation}
with $j\in\mathbb{Z}$ for even $\ell$ and $j\in\mathbb{Z}+1/2$ for odd $\ell$. The conservation of total angular momentum makes the angular part in Eq.~\eqref{HFloquet} diagonal, and the coupled differential equations governing the $n$th Floquet spinor component are 
\begin{subequations}\label{coupledEqs} \begin{align} &E^{m}_{j}\phi^{n}_{m,\uparrow,j}(r)\mathsmaller{=}\mathscr{L}^{\mathsmaller{-}}_{\alpha_n} \phi^{n}_{m,\downarrow,j}(r)\mathsmaller{+} M_{n_\mathsmaller{+}}\phi^{n}_{m,\uparrow,j}(r)\label{r1}\\ &{}^{+}\frac{\tilde{A}(r)}{2}\big[e^{i(\tau\mathsmaller{+}1)\theta}(1\mathsmaller{-}\tau)\phi^{n\mathsmaller{-}1}_{m,\downarrow,j}(r)\nonumber{\scriptstyle +}e^{\mathsmaller{-}i(\tau\mathsmaller{-}1)\theta}(1\mathsmaller{+}\tau)\phi^{n\mathsmaller{+}1}_{m,\downarrow,j}(r)\big]\nonumber\\ &E^{m}_{j}\phi^{n}_{m,\downarrow,j}(r)\mathsmaller{=}\mathscr{L}^{\mathsmaller{+}}_{\alpha_n\mathsmaller{-}1}\phi^{n}_{m,\uparrow,j}(r)\mathsmaller{+}M_{n_\mathsmaller{-}}\phi^{n}_{m,\downarrow,j}(r)\label{r2}\\ &{}^{+}\frac{\tilde{A}(r)}{2}\big[e^{i(\tau\mathsmaller{-}1)\theta}(1\mathsmaller{+}\tau)\phi^{n\mathsmaller{-}1}_{m,\uparrow,j}(r){\scriptstyle +}e^{\mathsmaller{-}i(\tau\mathsmaller{+}1)\theta}(1\mathsmaller{-}\tau)\phi^{n\mathsmaller{+}1}_{m,\uparrow,j}(r) \big],\nonumber \end{align} \end{subequations}
where $\alpha_n=j-(n-\frac{1}{2})(\ell+\tau)+\frac{1}{2}$, $M_{n_\pm}=\pm M+(n-\tfrac{1}{2})\hbar\Omega$, $E^{m}_{j}=\left(\epsilon_{m,j}-\hbar\Omega/2\right)$, and $\mathscr{L}^{\pm}_{\nu}=-i\hbar v_{\rm F}(\partial_r\mp \nu/r)$. Thus, total–angular–momentum conservation breaks $\mathcal{H}_{\rm F}$ into $j$ sectors, each governed solely by Eq.~\eqref{coupledEqs}, reducing it to an effectively $1$D eigenvalue problem.

\subsection{Effective Description of the VLB-driven System}

The driving protocol controls how a VLB imprints structure on the material. In the off–resonant case $\hbar\Omega\gg W$ (with $W$ the material’s bandwidth), absorption and emission are virtual and the conduction/valence bands remain effectively decoupled. The drive appears as an optical mass $M_{\rm opt}(r,\eta)\propto \tilde{A}^2(r)\sin\eta$ that mirrors the VLB envelope $A(r)$ [Fig.~\ref{Fig1}(b)], becoming maximal (minimal) for circular (linear) polarization. Because these effects stem from virtual absorption–emission processes, no net OAM is transferred and the leading response is OAM independent~\cite{Lauren2025}.

By contrast, in the resonant window $2M<\hbar\Omega<W$ Floquet replicas overlap and hybridize, opening light-induced gaps at avoided crossings. Near the one-photon resonance $\hbar\Omega/2$, a two-sector reduction (zero/one photon) of Eq.~\eqref{HFloquet} given by,
\begin{equation}\label{effectiveH}
H_{\rm F}(\mathbf r)=H_{\rm D}(\mathbf r)\sigma_{0}\alpha_{0}
+\frac{\hbar\Omega\sigma_0}{2}\alpha_z+\sum_{\gamma}\mathcal{H}_{\gamma}(\mathbf r)\alpha_{\gamma}
\end{equation}
captures the dynamics. Here, the Pauli matrices $\alpha$ act on the $n'=\{1,0\}$ subspace, and the quasienergy is given by $E_{m}=\epsilon_{m}-\hbar\Omega/2$. In this effective description the VLB’s OAM $\ell$ and polarization $\eta$ enter explicitly via the off-diagonal couplings $\mathcal{H}_{\gamma}$ [Eq.~\eqref{Hpm}], imprinting the beam’s structure into the quasienergy spectrum and eigenstates. Moreover, under weak coupling $g\equiv ev_{\rm F}A_0/(\hbar\Omega)\ll1$ additional anticrossings remain negligible and Eq.~\eqref{effectiveH} faithfully captures the driven system’s properties~\cite{graphene-top-ins,gaps-floq-graphene,OnePh1,OnePh2,OnePh3}. The remainder of the paper focuses on this effective description.

For CP the $n'=\{1,0\}$-projection of the total angular momentum operator in Eq.~\eqref{Jz2} takes the form 
\begin{equation}\label{Jz}
J^{\rm F}_{z}=\mathcal{L}_z\sigma_0\alpha_0+\frac{\sigma_z\alpha_0}{2}+(\ell+\tau)\frac{\sigma_0\alpha_z}{2}\;.
\end{equation}
Based on the $J^{\rm F}_{z}$ conservation, the states of the effective Hamiltonian become
\begin{eqnarray}\label{effective_states}
\hat{\Phi}^{\rm T}_{m,j}(\bm r)=\big[e^{-i(\ell+\tau+1)\theta}\phi^{1}_{m,\uparrow}(r), e^{-i(\ell+\tau)\theta}\phi^{1}_{m,\downarrow}(r),\nonumber\\ e^{-i\theta}\phi^{0}_{m,\uparrow}(r), \phi^{0}_{m,\downarrow}(r)\big]e^{i\ell_e\theta}\;,
\end{eqnarray}
and similar to the full Floquet case, they satisfy $J_z^{\rm F}\hat{\Phi}_{m,j}(\mathbf r)=j\hat{\Phi}_{m,j}(\mathbf r)= [\ell_e-(\ell+\tau+1)/2]\hat{\Phi}_{m,j}(\mathbf r)$ [Eq.~\eqref{JphiValue}]. Moreover, in the gauge of the Hamiltonian in Eq.~\eqref{effectiveH} the corresponding effective particle-hole operator reads  
\begin{equation}\label{pheff}
  P=\sigma_x\,\alpha_y\,K\;.
\end{equation}
Thus, if $H_{\rm F}\hat{\Phi}_{m}=E_m\hat{\Phi}_{m}$, then $H_{\rm F}(P\hat{\Phi}_{m})=-E_m(P\hat{\Phi}_{m})$. Equivalently, writing $P\hat{\Phi}_{m}= e^{i\delta}\hat{\Phi}_{\bar m}$ ($\delta\in\mathbb{R}$) yields $E_{\bar m}=-E_m$, pairing each state with a partner reflected about $E=0$ (i.e., about $\hbar\Omega/2$ in $\epsilon$). In addition, $P$ anticommutes with the effective total–angular–momentum operator, $\{J_z^{\rm F},P\}=0$. When $J_z^{\rm F}$ is conserved, $P$ maps states with $j\!\to\!-j$, {\it i.e.} if $J_z^{\rm F}\hat{\Phi}_{m,j}=j\hat{\Phi}_{m,j}$, then $J_z^{\rm F}(P\hat{\Phi}_{m,j})=-j(P\hat{\Phi}_{m,j})$, so $P\hat{\Phi}_{m,j}= e^{i\delta} \hat{\Phi}_{\bar m,-j}$. 

These symmetries manifest in the spectrum and in the system’s response to perturbations, as shown in the next sections. 
%


\section{Quasienergy Spectrum and States of VLB-driven System}\label{Sec3}
\begin{figure}[ht!]
  \centering
  \includegraphics[width=0.5\textwidth]{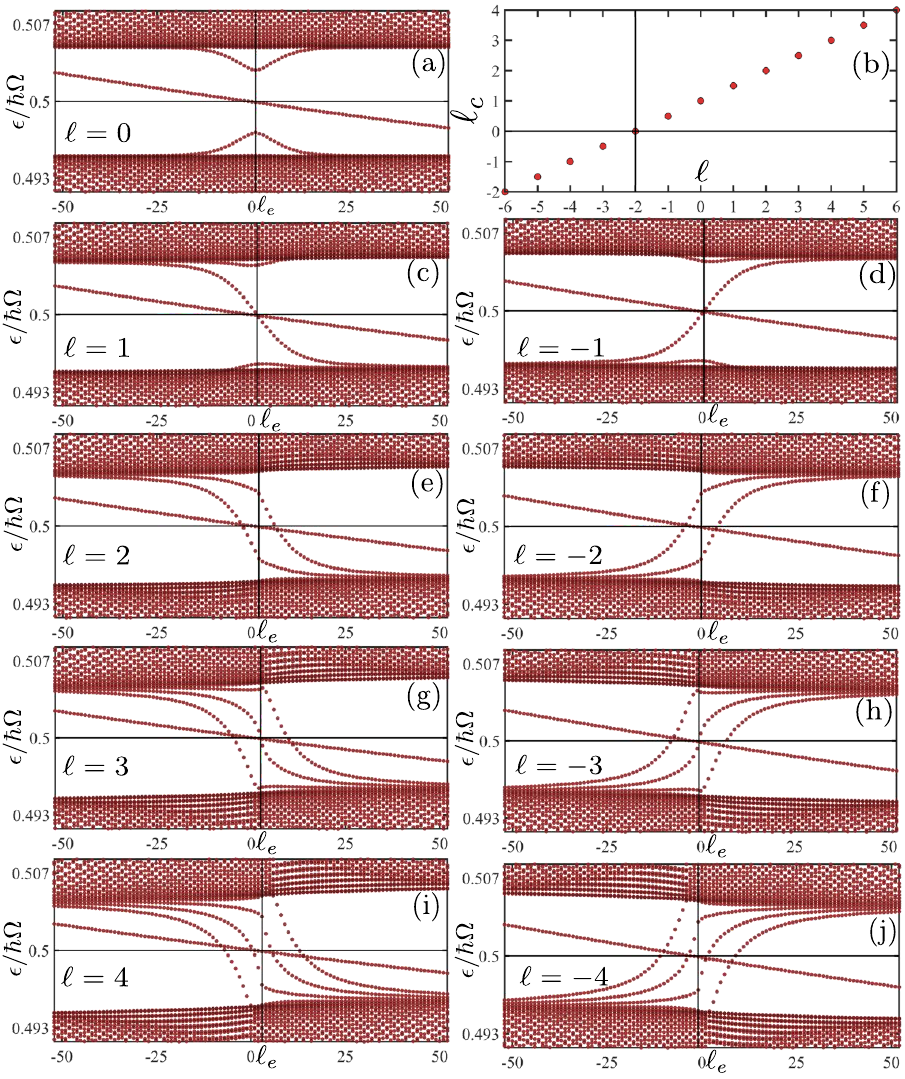}
  \caption{Panels (a), (c)–(j) show the quasienergy spectrum of a massive Dirac-like material with gap $2M$ under irradiation by a VLB. The system parameters are $\hbar\Omega=2.06M$, $k_{\rm F}\xi=10$, $\Delta_0=0.01M$, $\Delta(r)=\Delta_0\tanh(r/\xi)$, $R=10\xi$, and $v_{\rm F}/\Omega=1$\AA. The VLB's OAM, $\ell$, is indicated in each panel. The vertical line marks the particle–hole inversion center $\ell_c$, whose dependence on $\ell$ is shown in panel (b). Note that the vertical axis is $\epsilon=E+\hbar\Omega/2$.}
  \label{Fig2}
\end{figure}


For a general CP handedness $\tau$ the quasienergy spectrum is found by numerically diagonalizing the four linearly coupled equations given in Eq.~\eqref{coupledEqs} with $n=\{1,0\}$. The massive chiral Dirac fermions couple strongly to the $\tau$–handed CP light and only weakly to the opposite helicity $\bar{\tau}=-\tau$ due to chirality selection~\cite{Dch1,Dch2,Dh3}. Consequently, in the weak–coupling limit, the ratio of the $\bar{\tau}$ to $\tau$ dynamical gaps is $\widetilde{m}/(\hbar\Omega)$, $\widetilde{m}=(\hbar\Omega/2-M)$, so near resonance $\hbar\Omega\gtrsim 2M$ the $\bar{\tau}$ gap is negligible, scaling as $\sim g\,\widetilde{m}$. We set $\tau=1$, and the equations governing the system are explicitly given by 
\begin{align}\label{coupledEqsCP}
&\mathscr{L}^{-}_{\alpha_1} \phi^{1}_{m,\downarrow,j}(r)\mathsmaller{+} (\hbar\Omega/2\mathsmaller{-}\tilde{m})\phi^{1}_{m,\uparrow,j}(r)\mathsmaller{=}\,E^{m}_{j}\phi^{1}_{m,\uparrow,j}(r)\nonumber\\
&\mathscr{L}^{+}_{\alpha_1-1}\phi^{1}_{m,\uparrow,j}(r)\mathsmaller{+}\tilde{m} \phi^{1}_{m,\downarrow,j}(r)\mathsmaller{+}\tilde{A}(r)\phi^{0}_{m,\uparrow,j}(r)\mathsmaller{=}\,E^{m}_{j}\phi^{1}_{m,\downarrow,j}(r)\nonumber\\
&\mathscr{L}^{-}_{\alpha_0} \phi^{0}_{m,\downarrow,j}(r)\mathsmaller{-}\tilde{m} \phi^{0}_{m,\uparrow,j}(r)\mathsmaller{+}\tilde{A}(r)\phi^{1}_{m,\downarrow,j}(r)\mathsmaller{=}\,E^{m}_{j}\phi^{0}_{m,\uparrow,j}(r)\nonumber\\
&\mathscr{L}^{+}_{\alpha_0-1}\phi^{0}_{m,\uparrow,j}(r)\mathsmaller{-}(\hbar\Omega/2\mathsmaller{-}\tilde{m})\phi^{0}_{m,\downarrow,j}(r)\mathsmaller{=}\,E^{m}_{j}\phi^{0}_{m,\downarrow,j}(r)
\end{align}
where $\mathscr{L}$, $\alpha_n$, and $E^{m}_{j}$ are given in Eq.~\eqref{coupledEqs}.

\subsection{Fermion-doubling-free Treatment of the Driven System}~\label{doubling}
To obtain the quasienergy spectrum we discretize the radial coordinate and recast the coupled differential equations in each conserved angular-momentum channel $j$ into a matrix eigenvalue problem. A naive discretization of the first-order radial derivatives entering the operators $\mathscr{L}^{\pm}_{\alpha}$ using symmetric (central) finite-difference quotients generically suffers from the fermion-doubling problem~\cite{Stacey1982,szafran2019fd}. Mathematically, this originates from the replacement
$
\partial_r f(r)\approx[f(r+h)-f(r-h)]/{2h}+O(h^2),
$
for which the resulting lattice kinetic operator is invariant under the transformation $f_i \rightarrow (-1)^i f_i$ on the radial mesh $r_i=ih$. This even--odd (sublattice) symmetry allows additional low-energy lattice modes that have no continuum counterpart, appearing numerically as spurious eigenstates with rapidly oscillating radial profiles~\cite{szafran2019fd,Stacey1982}. Because the doubling stems from the discretization of the kinetic terms~\cite{Stacey1982,PhysRevC.106.L051303,szafran2019fd}, it is not removed by a finite Dirac mass $M$ (a local term)~\cite{szafran2019fd,PhysRevC.106.L051303}; likewise, the Floquet hybridization between photon sectors $n=\{1,0\}$ mediated by $\tilde{A}(r)$ is local in real space and therefore does not cure a discretization artifact rooted in the lattice representation of derivatives~\cite{Stacey1982,PhysRevC.106.L051303}.

To eliminate these spurious solutions, we adopt the asymmetric finite-difference scheme introduced in Ref.~\cite{PhysRevC.106.L051303}. In this approach, the central derivative is replaced by $q$-forward and -backward finite differences. For $q$$=$$3$, $\partial_r f(r_i)\approx[-3f_i+4f_{i+1}-f_{i+2}]/(2h)+O(h^2)$ (forward) or $\partial_r f(r_i)\approx[3f_i-4f_{i-1}+f_{i-2}]/(2h)+O(h^2)$ (backward), which are applied in a complementary way to the coupled spinor components in Eq.~(\ref{coupledEqsCP}) so that the Hamiltonian remains Hermitian~\cite{PhysRevC.106.L051303}. This procedure removes the even--odd mesh symmetry responsible for doubling and yields a one-to-one correspondence between lattice and continuum eigenstates within each $j$ sector, without introducing Wilson-type terms or modifying the continuum Hamiltonian~\cite{PhysRevC.106.L051303,szafran2019fd}. The resulting discretization is equivalent to the Stacey construction and to what is commonly referred to as the tangent-fermion method~\cite{Stacey1982,Beenakker2023TangentFermions}, which, as shown in Ref.~\cite{Beenakker2023TangentFermions}, preserves the topological properties of the underlying Dirac Hamiltonian.
\subsection{Spectral and Real-Space Signatures of Vortex and Edge States}
\begin{figure*}[ht!] 
  \centering
  \includegraphics[width=1\textwidth]{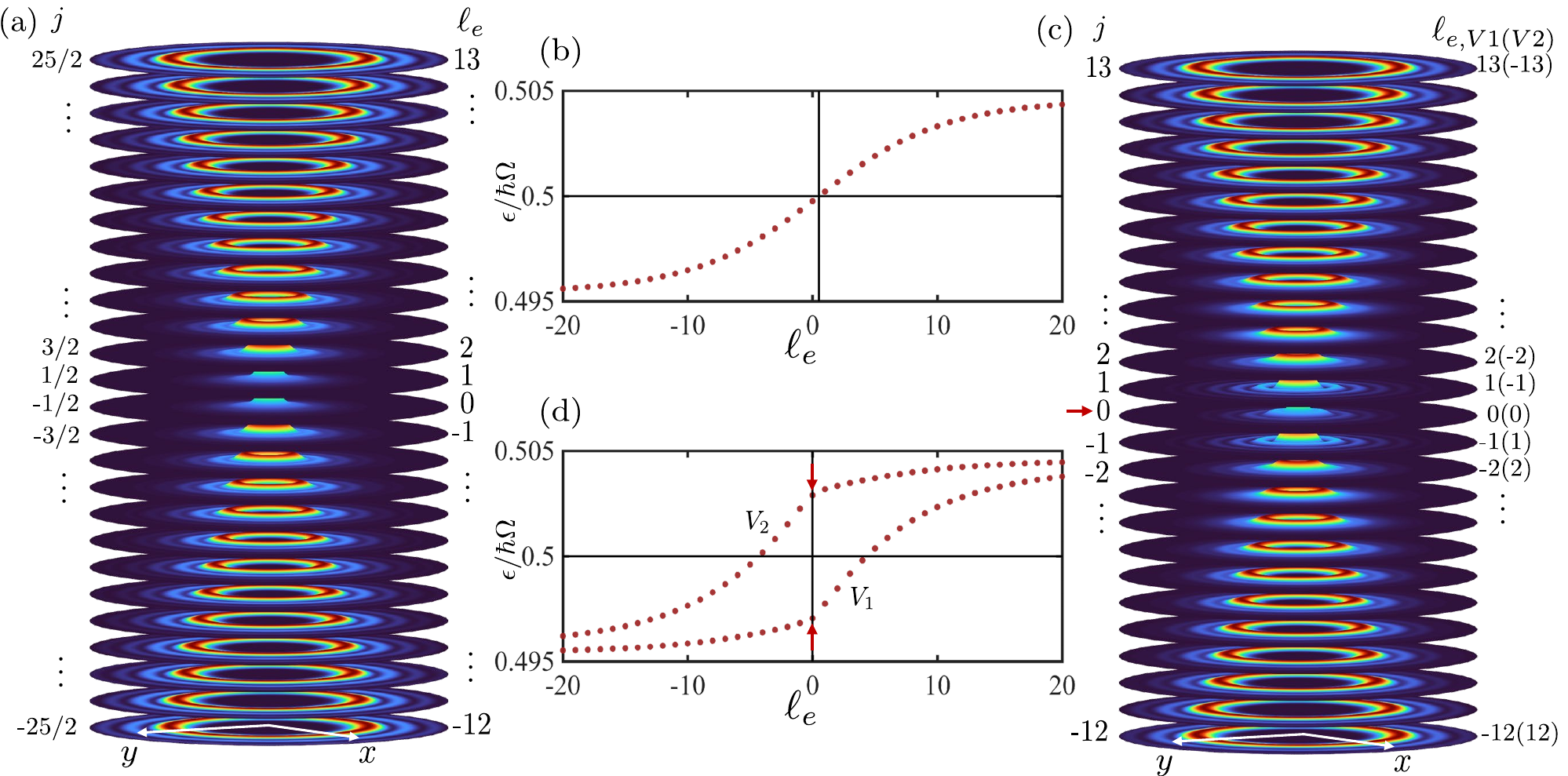}
\caption{(a) Real-space probability density of vortex states for the system of Fig.~\ref{Fig2}(d). Each disk corresponds to a state labeled by total angular momentum $j$ and orbital angular momentum $\ell_e$, ordered by increasing $j$ from bottom to top. The disk diameter is one quarter of the sample size, and the color scale runs from dark blue (low density) to dark red (high density).
(b) Quasienergy dispersion of the vortex branch associated with the states in (a).
(c) Real-space probability density of vortex states belonging to two vortex branches ($V_{1,2}$) for the system of Fig.~\ref{Fig2}(f).
(d) Corresponding quasienergy dispersion of the two vortex branches. The arrows indicate that the most confined states are not the ones closest to $E=0$, $\epsilon=\hbar\Omega/2$, [see the arrow in panel (c)]. For clarity, edge states and bulk bands are omitted in panels (b) and (d).}
  \label{Fig3}
\end{figure*}
The spectra and eigenstates are obtained numerically using the fermion-doubling–free discretization of Sec.~\ref{doubling}. Fig.~\ref{Fig2}(a),(c)–(j) shows the spectrum near the first Floquet Brillouin-zone edge $\hbar\Omega/2$, for circularly polarized VLBs ($\tau=1$) at several OAM values $\ell$, as a function of the orbital angular momentum $\ell_e$. Near the one-photon resonance, the quasienergy spectrum is characterized by three distinct types of states. The hybridization between Floquet states at $\hbar\Omega/2$ opens a dynamical gap of order $\sim e\hbar v_{\rm F}E_0/(\hbar\Omega)$~\cite{graphene-top-ins}. States above or below this gap are referred to as bulk states. In addition to bulk, there are two types of chiral in-gap states generated by the VLB’s SAM (CP handedness) $\tau$ and OAM $\ell$. The SAM leads to light-induced topological edge states~\cite{FloquetTI,FloquetTI2,graphene-top-ins}, while the OAM produces photoinduced, multiply quantized vortex states~\cite{Lauren2025}. The topological edge-state branch disperses linearly across the gap with a negative slope fixed by $\tau=1$, and thus appears in all panels (a),(c)–(j) of Fig.~\ref{Fig2}. On the other hand, for VLBs with $\ell\neq 0$, $|\ell|$ quantized vortex-state branches also appear with their slope near $\epsilon=\hbar\Omega/2$ set by ${\rm sgn}(-\ell)$~\cite{Lauren2025}.
\subsubsection{Vorticity}
Following the procedure presented in Ref.~\cite{Lauren2025}, Eq.~\eqref{effectiveH} maps to the Bogoliubov–de Gennes equations (BdGEs) for an $s$-wave superconductor (or superfluid) threaded by an $\ell$-flux–quanta~\cite{multiply,Caroli,ScBook,SCMVE2}, resulting in  
\begin{equation}\label{BdG1}
\left\{\frac{\sigma_z}{2m^*}\left[(\hbar k_{\rm F})^2\mathsmaller{-}{\bm p}^2\right]\mathsmaller{-}\Delta(r)\left[e^{i\ell\theta}\sigma_+\mathsmaller{+}e^{-i\ell\theta}\sigma_-\right]\right\}\widetilde{\Psi}=\mathscr{E}\widetilde{\Psi},
\end{equation}
where $m^{*}\equiv \hbar\Omega/(2v^2_{\rm F})$ plays the role of the effective mass in the BdGEs, $\widetilde{m}=\hbar^2 k_{\rm F}^2/(2m^{*})$, $\Delta(r)=\tilde{A}(r)$ serves as the (position-dependent) pairing amplitude, $\mathscr{E}=E^{m}_{j}$, and $\widetilde{\Psi}^{\rm T}=[\phi^{0}_{m,\uparrow}(\bm r), \phi^{1}_{m,\downarrow}(\bm r)]$. Notice that this equation can be cast in the form $H={\bm h}\cdot {\bm \sigma}$, with ${\bm h}$ defined as ${\bm h}=[-\Delta(r)\cos(\ell\theta),\Delta(r)\sin(\ell\theta),(\hbar^2 k_{\rm F}^2-{\bm p}^2)/(2m^{*})]$. The unit vector $\hat{\bm h}={\bm h}/|\bm h|$ is parameterized on the Bloch sphere by $\hat{\bm h}=[\sin(\Theta)\cos(\Phi),\sin(\Theta)\sin(\Phi),\cos(\Theta)]$. Since $h_x+ih_y=-\Delta(r)e^{-i\ell\theta}$, the Bloch-vector azimuth is locked to the real-space azimuthal angle, $\Phi(\theta)=-\ell\theta+\pi$. The polar angle follows $\tan\Theta(r)=\Delta(r)/\xi_k(r)$ with $\xi_k(r)=(\hbar^2 k_{\rm F}^2-{\bm p}^2)/(2m^{*})$ 
and $\widetilde m$ fixed, so at each radius the latitude is set smoothly by the VLB envelope together with a local kinetic-energy offset (the sign of $\xi_k$ placing the Bloch vector above or below the equator). By holding $r$ fixed, the Bloch vector is pinned to a constant latitude $\Theta(r)$. Then, when encircling the origin in real space ($\theta:0\to2\pi$), the azimuth advances by $\Delta\Phi=-2\pi\ell$, winding the Bloch vector $|\ell|$ times. The VLB-driven system is therefore characterized by a vorticity $w=\Delta\Phi/(2\pi)=-\ell$~\cite{vorticity}.
\subsubsection{Vortex Branches Dispersion}
To analyze the spectrum, we review relevant results from Ref.~\cite{Lauren2025} and contextualize them in the particle-hole framework. We first fix total angular momentum and separate the angular dependence with the ansatz $\widetilde{\Psi}^{\rm T}_{m,j}(\bm r)=[e^{-i\theta}\phi^{0}_{m,\uparrow}(r),e^{-i(\ell+1)\theta}\phi^{1}_{m,\downarrow}(r)]$. This reduction yields a purely radial equation,
\begin{equation}\label{mappedH}
- \frac{\hbar^2\sigma_z}{2m^{*}}\left[\partial_r^2+ \frac{\partial_r}{r} - \frac{\alpha_\ell^2}{r^2} + k^2_{\rm F} -\sigma_z\frac{\beta}{r^2}\right]\hat{\Phi} + \sigma_x\Delta(r)\hat{\Phi} = \mathscr{E}\hat{\Phi},
\end{equation}
with $\hat{\Phi}^{\rm T}(r)=[\phi^{0}_{m,\uparrow}(r), \phi^{1}_{m,\downarrow}(r)]$, $\alpha_\ell=\sqrt{j^2+\ell^2}$, and $\beta=j\ell$. Equation~\eqref{mappedH} mirrors the BdG description of an $s$-wave SC/superfluid hosting an $\ell$-flux–quanta vortex and thus encodes both bulk and vortex features of the VLB-driven system~\cite{multiply,Caroli,ScBook,SCMVE2}. Leveraging this correspondence, one may follow the procedure in Ref.~\cite{multiply}. This technique exploits the small-$r$ behaviour and the asymptotic limit for Eq.~\eqref{mappedH} to obtain the vortex-branch quasienergies near the one-photon resonance. The VLB profile $\Delta(r)=\Delta_0\tanh(r/\xi)$ simplifies to the closed form
\begin{equation}\label{omegas}
\mathscr{E^{\rm V}}=-\omega_0\ell j + b\left(s+\frac{|\ell| -1}{2}\right)\tilde{\omega},
\end{equation}
with $s\in \mathbb{Z}$, $\omega_0=a \Delta_0/(k_{\rm F}\xi)$, $\tilde{\omega}=bk_{\rm F}/{\xi}$, and $b/a=\pi^3/[14\zeta(3)]\approx 1.8$. The quasienergies closest to $\mathscr{E^{\rm V}}=0$ (i.e., to the one-photon resonance) occur at
\begin{equation}\label{jns}
j_{s}=\frac{b}{a}\left(s+\frac{|\ell| -1}{2}\right)\frac{k_{\rm F}\xi}{\ell},
\end{equation}
and enforcing $|j|\lesssim k_{\rm F}\xi$ yields $-|\ell|+1/2\lesssim s\lesssim 1/2$, i.e., $s=0,-1,-2,\dots,1-|\ell|$. This counting demonstrates that the absolute value of the  vorticity $|\ell|$ determines the number of vortex-state branches. 

The vortex structure, together with particle-hole symmetry, leads to a natural classification depending on whether $|\ell|$ is even or odd. A specific value of $\ell=\ell_c$ identifies the particle-hole inversion center, as shown in Fig.~\ref{Fig2}(b). For even $|\ell|$, the spectrum contains $|\ell|/2$ pairs of vortex branches within the dynamical gap, with electron-hole symmetric states connected across branches. For odd $|\ell|$, there is a central vortex branch that disperses linearly as $\mathscr{E}^{\rm {V}_0}_{\ell_e}\approx-j\omega_0$ and contains electron-hole symmetric states. In addition there are $(|\ell|-1)/2$ pairs of vortex branches with even-$\ell$-like electron-hole pairs. 



\subsubsection{Real Space Distribution of Vortex States}
Within this unified picture, the vorticity fixes the number and directionality of the vortex branches, while particle-hole symmetry determines how states are paired in $(\epsilon,j)$. By contrast, the topological edge states are controlled only by the light polarization: since $\tau=1$ is fixed, their linear dispersion is unchanged across the quasienergy panels of Fig.~\ref{Fig2}. Since $j=\ell_e-\ell/2-1$ (for $\tau=1$) is integer for even $\ell$ and half integer for odd $\ell$, the value of $\ell_c$ is obtained by exploiting the particle-hole symmetry of the spectrum, according to Eq.~\eqref{omegas}. For integer $j$, $\ell_c$ corresponds to $\ell_e$ that makes $j=0$, and for half-integer $j$, $\ell_c$ interpolates between the two values of $j$ nearest to $0$, Fig.~\ref{Fig2}(b). Since this symmetry relates $(E,j)\rightarrow(-E,-j)$ [$E=\epsilon-\hbar\Omega/2$], the partner states differ only by a phase and therefore have identical real-space probability densities.  
Figures~\ref{Fig3}(a) and (c) show the real-space density of the vortex states for $\ell=-1$ and $-2$, respectively. Each slice in the stacked plot represents the probability density of a vortex state labeled by its $j$ (or $\ell_e$) across the vortex branches shown in Figs.~\ref{Fig3}(b) and (d). The vortex states are localized within the VLB-induced core and are characterized by a localization radius $R_j$ that increases with $|j|$. The plots clearly show that particle-hole partners have identical real-space distributions and therefore the same localization radius $R_j$.

This radial structure can be understood semiclassically from the mapped BdG problem in Eq.~\eqref{mappedH}. There, the VLB OAM $\ell$ plays the role of the vortex winding, and the angular-momentum terms controlled by $\alpha_\ell=\sqrt{j^2+\ell^2}$ and $\beta=j\ell$ determine the radial support of the spinor components, as in the $s$-wave SC/superfluid threaded by an $\ell$-flux–quanta~\cite{multiply}. The leading centrifugal scale is set by the $\alpha_\ell^2/r^2$ term, giving $R_j\sim \alpha_\ell/k_{\rm F}$ 
while $\beta$ mainly introduces an  asymmetry between the spinor components without changing this scale, consistent with Fig.~\ref{Fig3}.

An additional constraint on the relation between $j$ and the real-space structure of the vortex states follows from the integer--half-integer alternation of $j$ and the corresponding particle-hole symmetry center $\ell_c$ [Fig.~\ref{Fig2}(b)]. In particular, when $\ell=-1$, there is a single vortex branch ($j$ is half integer). The states closest to $E=0$ also have the smallest values of $|j|$ and the smallest localization radii, as seen in Figs.~\ref{Fig3}(a) and (b). 
By contrast, when $\ell=-2$, there are two vortex branches ($j$ is integer), and particle-hole symmetry connects states across different branches. Interestingly, the states with the smallest localization radii are not the ones closest to $E=0$, as illustrated in Figs.~\ref{Fig3}(c) and (d). Particle-hole symmetry also implies the presence of two states with $j=0$, while all other partners 
have identical $R_j$ in real space. In summary, for odd $|\ell|$ the central branch enforces a direct correspondence between proximity to $E=0$ and minimal localization radius, whereas for even $|\ell|$ the paired-branch structure breaks this correspondence and redistributes the most localized states away from $E=0$.

In both cases, states with smaller localization radii $R_j$ exhibit higher probability density because their spatial extent is more confined. This is visible in the stacked plots of Figs.~\ref{Fig3}(a) and (c). We note, however, that for visualization purposes the peak densities of these strongly localized states are partially truncated in the stacked representation. 

\subsubsection{Local Density of States of VLB Driven System}\label{LDOS}
A promising route for Floquet engineering is to use external reservoirs to stabilize the driven nonequilibrium steady state over time~\cite{flreview5,OcupationFTI1,inti24}. Coupling to bosonic or fermionic reservoirs introduces inelastic scattering channels that balance the energy injected by the drive and allow the system to reach a nonequilibrium steady state~\cite{Ocupation1}. In particular, when the driven system is coupled to a metallic lead through an intermediate resonant level, the lead provides an energy-selective channel, making it natural to describe the steady-state population in terms of an effective Floquet Fermi level, or quasienergy chemical potential, that separates predominantly occupied from predominantly empty Floquet states~\cite{Ocupation1,Ocupation2,Ocupation3,Ocupation4,Asmar2021,Asmar2020,Asmar2021_2}. The resulting nearly Fermi-like occupation is then smoothly smeared over an energy scale $\Gamma$~\cite{Ocupation1}. In our work, we adopt this phenomenological picture to describe quantities such as the LDOS and to apply the perturbative analysis presented below. Accordingly, the LDOS of the VLB-driven system is written as
\begin{align}\label{LSOS-clean}
\rho(\bm r,\epsilon_0)&=\sum_{m,j}g_{\Gamma}\left(E_{0}-E_{m,j}\right)\,\rho_{m,j}(\bm r)\;,\nonumber\\
\rho_{m,j}(\bm r)&= \big|\hat{\Phi}_{m,j}(\bm r)\big|^2\qquad\text{[see Eq.~\eqref{effective_states}]}\;, \nonumber\\
g_{\Gamma}(E_{0}-E_{m,j})&=\frac{1}{\sqrt{2\pi}\Gamma}\exp\!\left[-\frac{(E_{0}-E_{m,j})^2}{2\Gamma^2}\right]\;.
\end{align}
This expression is a broadened form of the standard spectral representation
$\rho(\bm r,\epsilon)=\sum_{m,j}\rho_{m,j}(\bm r)\,\delta(\epsilon-E_{m,j})$,
in which each discrete quasienergy peak at $E_{m,j}$ is replaced by a Gaussian of width $\Gamma$. Each Floquet eigenstate contributes through its local weight
$\rho_{m,j}(\bm r)=|\hat{\Phi}_{m,j}(\bm r)|^2$, while $g_\Gamma(E_{0}-E_{m,j})$ selects states within
$|E_{m,j}-E_{0}|\lesssim \Gamma$ of the probe quasienergy $E_{0}$. 
\begin{figure}[ht]
  \centering
  \includegraphics[width=0.5\textwidth]{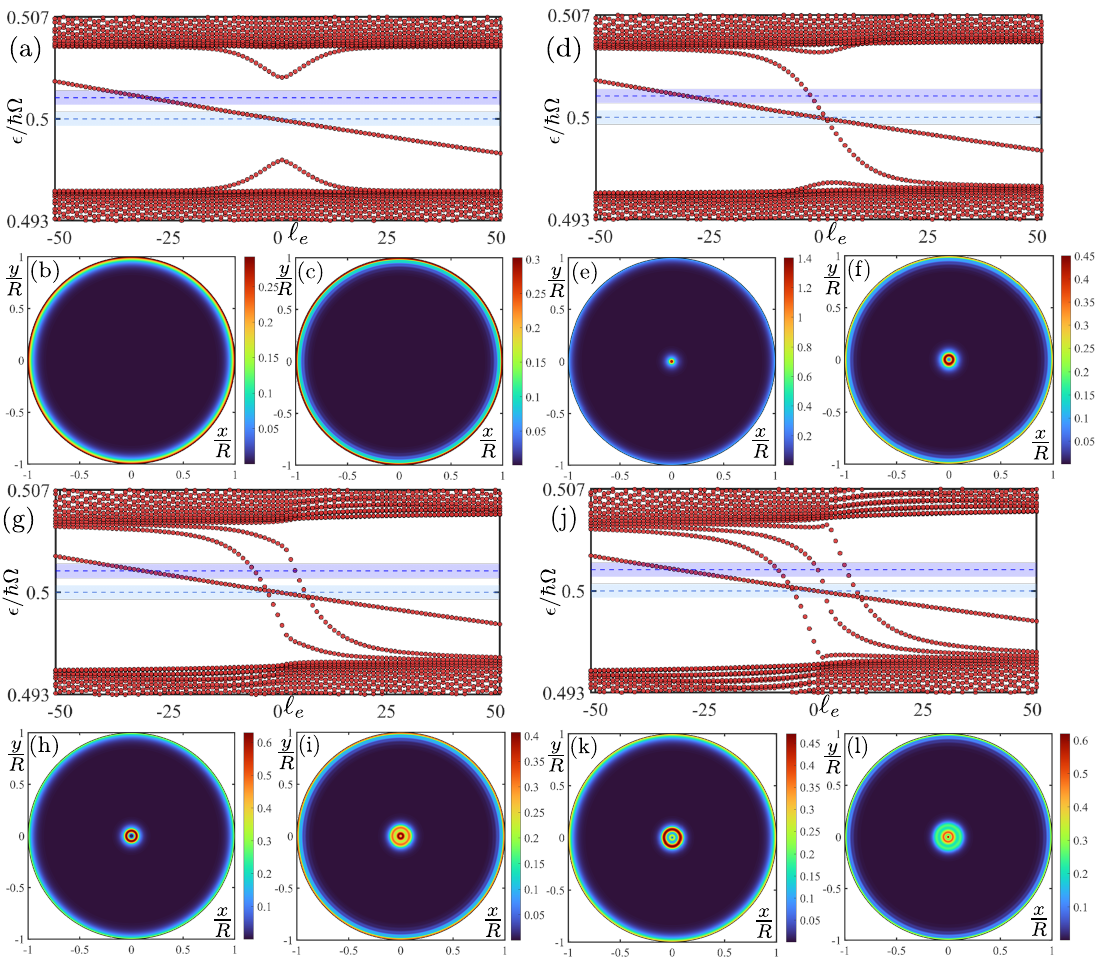}
\caption{LDOS of the VLB-driven system. The probe quasienergies $\epsilon_0=E_{0}+\hbar\Omega/2$ and the broadening $\Gamma$ are indicated in the quasienergy spectra shown for $\ell=0,1,2,$ and $3$ in panels (a), (d), (g), and (j), respectively; all other material and VLB parameters are the same as in Fig.~\ref{Fig2}. Panels (b), (e), (h), and (k) show the LDOS for $E_{0}=0$, while the remaining panels show the LDOS for $E_{0}=0.003M$. In all cases, $\Gamma=0.001M$.}
  \label{Fig4}
\end{figure}


In Fig.~\ref{Fig4} we show the quasienergy spectra and the corresponding LDOS for selected probe quasienergies $E_{0}$, a fixed reservoir-induced broadening $\Gamma$, and several values of the VLB OAM. For $\ell=0$ [Fig.~\ref{Fig4}(a)], one clearly observes the topologically protected edge states associated with a Floquet topological insulator, which emerges when a massive Dirac-like material is driven resonantly by CP light~\cite{FloquetTI}. These edge states appear prominently in the LDOS at the sample boundary [Figs.~\ref{Fig4}(b) and (c)]. Because of the finite broadening $\Gamma$, the LDOS receives contributions from several nearby edge states, producing spectral weight concentrated uniquely at the edge and highlighting their strong edge localization. Moreover, in the $\ell=0$ calculation we retain the VLB beam profile to make clear that no vortex states are present in this case, showing that vortex-state formation is induced specifically by the VLB’s OAM rather than by the beam shape itself.

For $\ell=1$, the LDOS at $E_{0}=0$ [Fig.~\ref{Fig4}(e)] is visibly more localized than the LDOS at $E_{0}=0.003M$ [Fig.~\ref{Fig4}(f)]. This follows from the fact that the states contributing to $\rho(\bm r,E_{0}=0)$ have smaller values of $j$ and therefore smaller localization radii, as shown in Fig.~\ref{Fig3}, whereas the states contributing to $\rho(\bm r,E_{0}=0.003M)$ are associated with larger $j$ and thus extend farther from the core. Comparing Figs.~\ref{Fig4}(e) and (f) with the individual-state densities in Fig.~\ref{Fig3} also clarifies the role of the induced broadening: the LDOS is not the profile of a single state, but a weighted superposition of several nearby states. Since, in this case, the contributing states have adjacent localization radii $R_j$, the resulting LDOS resembles a single broadened state with an effective broadened localization radius. In both Figs.~\ref{Fig4}(e) and (f), topological edge states are still present. However, in Fig.~\ref{Fig4}(e), the vortex states with the smallest radii produce a much larger LDOS 
than the corresponding contribution from edge states (note the color scale).
In Fig.~\ref{Fig4}(f), the vortex states are more extended with their LDOS contribution becoming comparable to that of the edge states.

For $\ell=2$ [Fig.~\ref{Fig4}(g)], the states contributing to $\rho(\bm r,E_{0}=0)$ [Fig.~\ref{Fig4}(h)] are particle-hole symmetric states drawn from both vortex branches. Particle-hole symmetry imposes $R_j=R_{-j}$, which leads to a single ring-like feature in the LDOS. 
By contrast, for $E_{0}=0.003M$ the contributing states are not particle-hole symmetric and span different localization radii, leading to the appearance of multiple high-density rings in Fig.~\ref{Fig4}(i). Notice that the high-density radii in the LDOS for $\ell=1$ are produced by states with consecutive values for their radii $R_{j}$, while for $\ell=2$ this is not always the case. For a general $E_{0}$ there is a large variation in the values of $R_{j}$, except for the case of $E_{0}=0$ where they satisfy  $R_j=R_{-j}$, where particle-hole symmetry manifests.    
In all panels [Figs.~\ref{Fig4}(g)–(i)], the edge states remain visible, with densities comparable to that of the vortex states.

Fig.~\ref{Fig4}(j) shows the case of $\ell=3$. Here, the LDOS for $E_{0}=0$ [Fig.~\ref{Fig4}(k)] receives contributions from both the central branch and the two outer branches. The central-branch states have the smallest localization radii $R_j$, while the outer-branch states extend to larger radii and contribute more strongly at larger distances from the core. As a result, $\rho(\bm r,E_{0}=0)$ exhibits an inner ring associated with the central-branch weights and a broader, more pronounced outer ring associated with the outer branches. In this case, only two rings are visible because the states contributing to the LDOS are particle-hole symmetric partners and therefore combine into the same radial structures. By contrast, for $\rho(\bm r,E_{0}=0.003M)$ [Fig.~\ref{Fig4}(l)], more than two regions of enhanced density appear, since the contributing states are not particle-hole symmetric partners. The innermost region arises from the central-branch states, while the outer regions originate from the weights of the states in the outer branches. The outermost ring is less intense than the middle one as it arises from the rightmost vortex branch, which contributes fewer states within the broadening window. In this case, the LDOS appears more spatially spread due to the wider range of vortex states included within the energy window set by $\Gamma$. Finally, in all panels for $\ell=3$, the topological edge states remain clearly visible and coexist with the vortex states, with their spectral weight localized at the sample boundary.


\section{Angular Momentum Mixing: Impurities and Deviations from Circular Polarization}\label{Sec4}

To examine the effect of imperfections on the pristine VLB-driven system, we assume, as in Sec.~\ref{LDOS}, that the driven system reaches a nonequilibrium steady state through coupling to external reservoirs. Within this framework, the unperturbed basis is given by the Floquet states of the ideal VLB-driven system, {\it i.e.} the states in Eq.~\eqref{effective_states}. Perturbations that break the conservation of $J^{\rm F}_{z}$, {\it i.e.} $[\hat{V}(\bm r),J^{\rm F}_{z}]\ne 0$ [Eq.~\ref{Jz}], modify both the quasienergy spectrum and the corresponding photon-dressed states. 
In this section, we focus on two sources of angular-momentum mixing: local defects modeled as impurities, and deviations of the driving beam from CP.

\subsection{Impurity Effects on Quasienergy Spectrum, States, and LDOS}\label{ImpurityPert}
We model an impurity potential in the equilibrium material by a Gaussian profile with spatial extent $a_0$, strength $V_{0}=V/(2\pi a^2_0)$, and position ${\bm r}_0=(r_0,\theta_0)$,
$
\mathcal{V}(\bm r)=V_{0}\exp\!\left[-|{\bm r}-{\bm r}_0|^2/(2a_0^2)\right]\sigma_0\;
$, Fig.~\ref{Fig1}(e).
We assume that the impurity potential is smooth on the lattice scale, $a_0\gg a$, where $a$ is the interatomic distance. Under this condition, it can be incorporated directly into the low-energy Dirac Hamiltonian as
$H_{\rm D,imp}=H_{\rm D}(\bm r)+\mathcal{V}(\bm r)$.

We further take the impurity to be static, so that the VLB drive does not induce any time dependence in $\mathcal{V}(\bm r)$. Consequently, the impurity potential is diagonal in Floquet space, entering the Floquet Hamiltonian of Eq.~\eqref{HFloquet} as $\mathcal{V}(\bm r)\delta_{n',n}$. In the effective Floquet Hamiltonian $H_{\rm F}(\bm r)$ [Eq.~\eqref{effectiveH}], this corresponds to a scalar potential term,
\begin{equation}\label{impurity}
  \mathcal{V}(\bm r)=V_{0}\exp\!\left[-\frac{|{\bm r}-{\bm r}_0|^2}{2 a_0^2}\right]\sigma_0\alpha_0\;.
\end{equation}

We treat the impurity as a perturbation to the Floquet states and its quasienergies [Eq.~\eqref{effective_states}]. The first-order correction to the quasienergy is
\begin{align}\label{epsilon_c}
  \Delta E^{m}_{j}=V_{0}\int\exp\!\left[-\frac{|{\bm r}-{\bm r}_0|^2}{2 a_0^2}\right]\rho_{m,j}(\bm r)\,rdrd\theta\;,
\end{align}
where $\Delta E^{m}_{j}$ is the first-order correction to $E^{m}_{j}$ in Eq.~\eqref{coupledEqsCP}, and $\rho_{m,j}(\bm r)=|\hat{\Phi}_{m,j}(\bm r)|^2$ is the local probability density of the unperturbed Floquet state $\hat{\Phi}_{m,j}(\bm r)$.

\begin{figure}[ht!]
  \centering
  \includegraphics[width=0.5\textwidth]{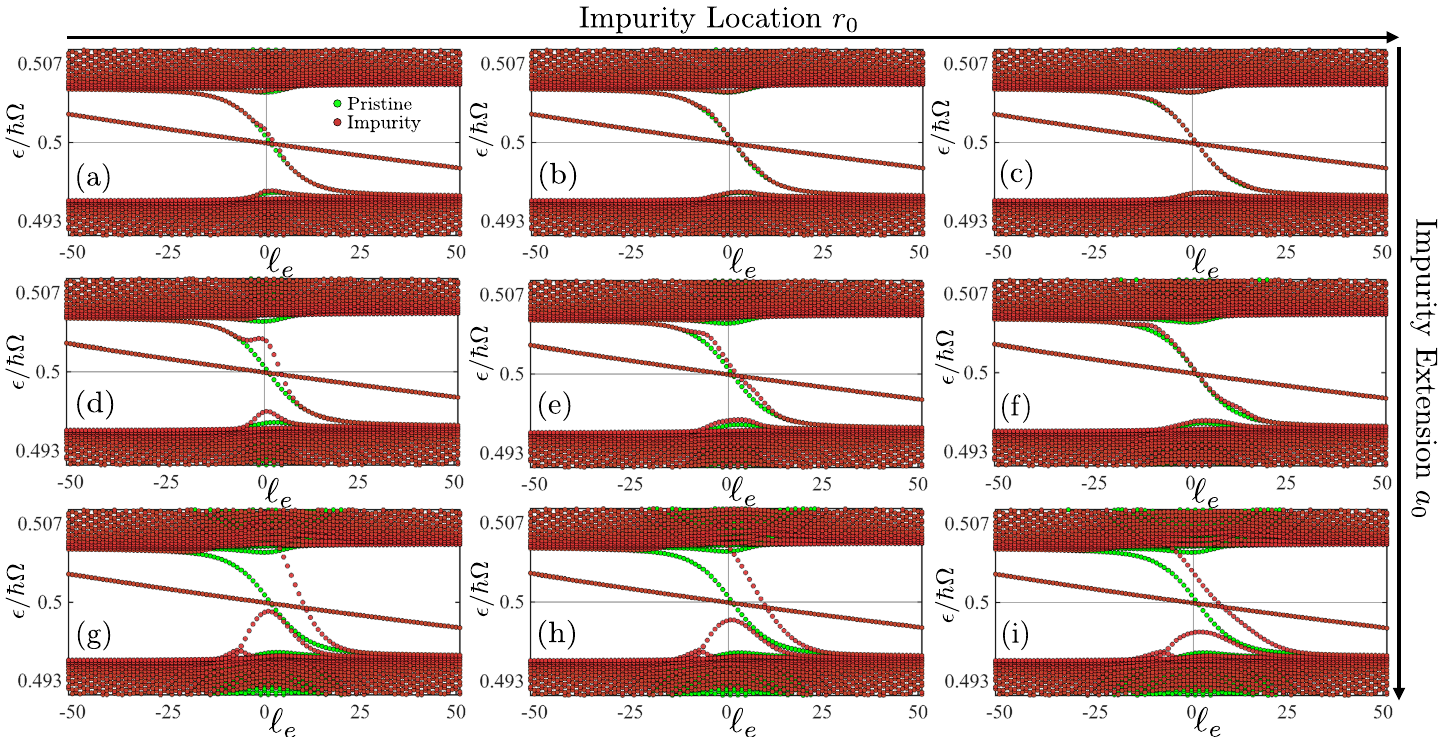}
  \caption{Perturbed quasienergy for an impurity placed along the $x$ axis at radial position $r_0$, with strength $V_0=0.85(2g)$, where $2g$ is the dynamical gap and $g\equiv ev_{\rm F}A_0/(\hbar\Omega)$~\cite{strong}. Red dots denote the perturbed spectrum, while green dots show the corresponding unperturbed spectrum. The impurity widths are $a_0/\xi=0.1$ in (a)--(c), $0.25$ in (d)--(f), and $0.75$ in (g)--(i). In each row, the impurity positions are $r_0/\xi=0.25,\,0.75,$ and $1.25$ from left to right. All other material and VLB parameters are the same as in Fig.~\ref{Fig2}(c).}
  \label{Fig5}
\end{figure}
To gain physical insight into how the impurity shifts the in-gap quasienergies, we use the fact that vortex and edge states are strongly localized around a characteristic radius, as shown in Figs.~\ref{Fig3} and \ref{Fig4}. Thus, for a localized state we approximate its density by
$\rho_{m,j}(\bm r)\approx \rho_{m,j}(\bm R^{m}_{j})\delta(\bm r-\bm R^{m}_{j})$,
where $\bm R^{m}_{j}=(R^{m}_{j},\theta)$. Here, $R^{m}_{j}$ is the localization radius of the state with total angular momentum $j$ in the $m^{\rm th}$ Floquet sideband. This approximation neglects the oscillatory decay of the wave function away from its dominant radial support. Within this localized-state approximation, Eq.~\eqref{epsilon_c} gives
\begin{align}\label{epsilon_c_approx}
  \Delta E^{m}_{j}\approx V_{0}\exp\!\left[-\frac{|{\bm R}^{m}_{j}-{\bm r}_0|^2}{2 a_0^2}\right]\rho_{m,j}(\bm R^m_{j})\;. 
\end{align}
Since the impurity strength sets the overall magnitude of the perturbation, the quasienergy correction $\Delta E^{m}_{j}$ scales linearly with $V_0$. 
Notice that, the spatial overlap between the impurity profile and the localized states directly determines the modifications to the quasienergy spectrum displayed in Fig.~\ref{Fig5}. Recalling that the particle-hole symmetry of the unperturbed system is reflected in the real-space distributions of the states, {\it i.e.} $\rho_{m,j}(\bm R^{m}_{j})=\rho_{\bar m,-j}(\bm R^{\bar m}_{-j})$ as shown in Fig.~\ref{Fig3}, it follows that particle-hole partner states experience identical quasienergy shifts, $\Delta E^{m}_{j}=\Delta E^{\bar m}_{-j}$. 
\begin{figure*}[ht!] 
  \centering
  \includegraphics[width=1\textwidth]{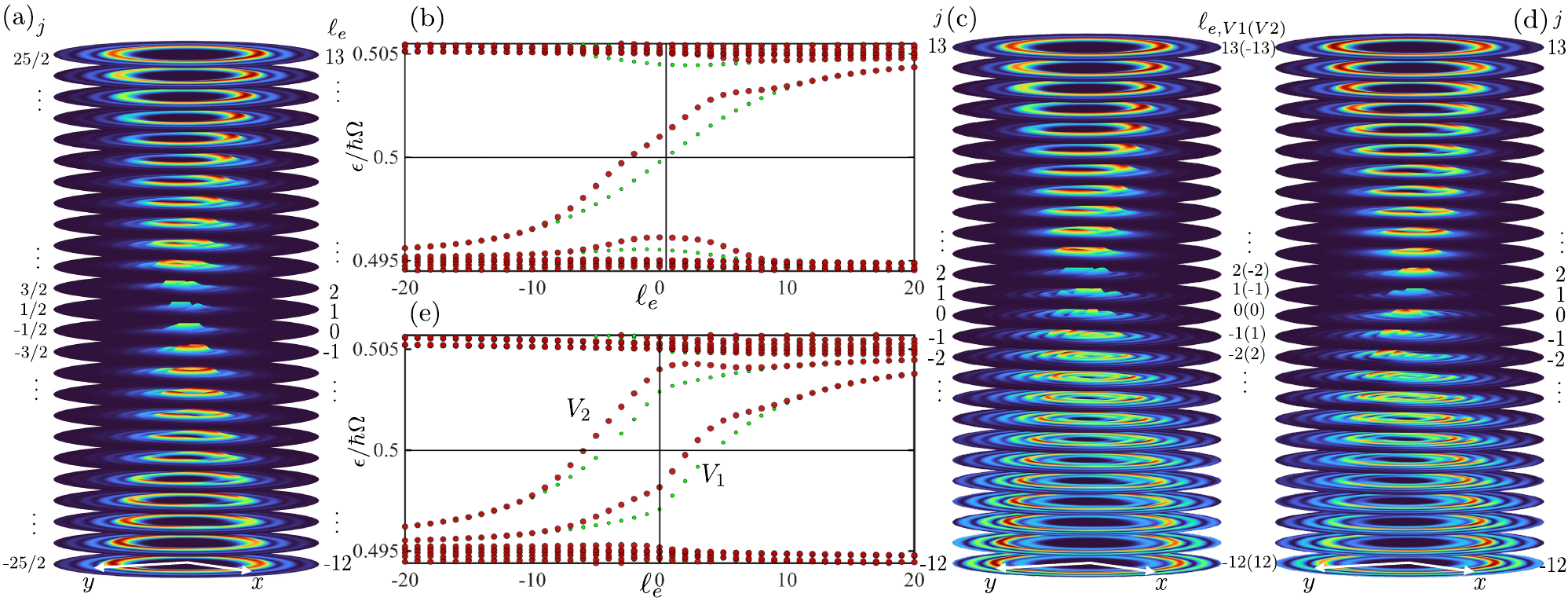}
\caption{(a) Real-space probability density of the vortex states of Fig.~\ref{Fig2}(d) in the presence of a Gaussian impurity with $a_0/\xi=0.25$ located along the $x$ axis at $r_0/\xi=0.5$. Each disk corresponds to a state labeled by $j$ and $\ell_e$, ordered by increasing $j$ from bottom to top; the disk diameter is one quarter of the sample size, and the color scale runs from dark blue (low density) to dark red (high density). (b) and (e) Corresponding vortex-branch quasienergy dispersion, with red dots denoting the impurity-corrected spectrum and green dots the unperturbed one. (c) Real-space probability density of vortex states in branches $V_1$ and $V_2$ for the system of Fig.~\ref{Fig2}(f), with the same impurity configuration as in (a). (d) Corresponding quasienergy dispersion of the two vortex branches. In panels (b) and (e), the edge states are not shown for clarity.}
  \label{Fig6}
\end{figure*}

We divide the analysis of Fig.~\ref{Fig5} according to the impurity extent by considering short-range impurities (highly localized) [Figs.~\ref{Fig5}(a)–(c)], medium-range impurities [Figs.~\ref{Fig5}(d)–(f)], and extended ``clusters of impurities'' [Figs.~\ref{Fig5}(g)–(i)]~\cite{clusters}. For short-range impurities, the spatial range of the impurity overlaps with only a small number of localized states. Since the localization radius of the vortex states increases with $|j|$, impurities positioned close to the vortex core, $r_0/\xi=0.25$, primarily affect the states with the smallest values of $|j|$, as shown in Fig.~\ref{Fig5}(a). As the impurity position is moved outward to $r_0/\xi=0.75$ [Fig.~\ref{Fig5}(b)], 
it overlaps more with states of larger $|j|$, making them the most perturbed states. Finally, when the impurity is placed at $r_0/\xi=1.25$ [Fig.~\ref{Fig5}(c)], nearly all vortex states lie outside the effective range of the impurity, so the quasienergy corrections become negligible and the spectrum approaches that of the pristine system. 

For medium-range impurities [Figs.~\ref{Fig5}(d)–(f)], the larger impurity extent produces a substantially stronger modification, since the impurity overlaps with a much larger number of localized states. This is already evident for $r_0/\xi=0.25$ [Fig.~\ref{Fig5}(d)], where many low-$|j|$ states acquire noticeable quasienergy shifts. As the impurity is moved outward to $r_0/\xi=0.75$ [Fig.~\ref{Fig5}(e)], the perturbation extends over a broader range of vortex states, leading to pronounced corrections for states with larger values of $|j|$. Unlike the short-range case, the effect of the impurity remains significant even at $r_0/\xi=1.25$ [Fig.~\ref{Fig5}(f)], where its larger spatial extent still overlaps with vortex states localized farther from the core, particularly those with higher $|j|$.

For extended ``clusters of impurities'' [Figs.~\ref{Fig5}(g)–(i)], the impurity profile overlaps with most of the vortex states. As a result, the impurity induces nearly uniform quasienergy corrections across the vortex branches, effectively behaving as an approximately constant background potential. This manifests itself as an almost rigid upward shift of the vortex-state quasienergies in Figs.~\ref{Fig5}(g)–(i). As the center of the impurity cluster is moved farther away from the vortex core, the overall magnitude of this shift gradually decreases due to the reduced spatial overlap between the impurity profile and the localized vortex states.

The edge-state branch is comparatively weakly affected by the impurity configurations considered here. For impurities centered near the vortex core, the overlap with edge-localized states is small, and the edge dispersion remains nearly unchanged. When the impurity profile overlaps appreciably with the boundary, its main effect is a nearly uniform quasienergy shift of the edge branch, since the edge states sample a similar scalar potential along the sample edge. Within the smooth scalar-potential perturbations considered here, the chiral edge branch is not gapped out as long as the dynamical gap remains open, consistent with its topological origin.

After obtaining the diagonal quasienergy correction $\Delta E_j^{m}$ [Eq.~\eqref{epsilon_c}], we next consider the impurity-induced modification of the Floquet states themselves. Treating the impurity perturbatively in the basis of the pristine Floquet states $\{\hat{\Phi}_{m,j}(\bm r)\}$, the coupling between the states $(m,j)$ and $(m',j')$ is determined by the matrix element
\begin{align}
\mathcal{V}^{m'j'}_{mj}
=\int d^2r\;\hat{\Phi}^{\dagger}_{m',j'}(\bm r)\,\mathcal{V}(\bm r)\,\hat{\Phi}_{m,j}(\bm r),
\end{align}
such that the impurity-corrected Floquet state is given, within first-order perturbation theory, by
\begin{align}\label{correctedstate}
\hat{\Phi}^{(c)}_{m,j}(\bm r)
=\mathcal{N}_{m,j}\left[\hat{\Phi}_{m,j}(\bm r)+
\sum_{\substack{j'\neq j\\m,m'}}
\frac{\mathcal{V}^{m'j'}_{mj}}
{E^{m}_{j}-E^{m'}_{j'}}\,
\hat{\Phi}_{m',j'}(\bm r)\right],
\end{align}
where $\mathcal{N}_{m,j}$ is a normalization factor. 

\begin{figure}[ht!] 
  \centering
  \includegraphics[width=0.5\textwidth]{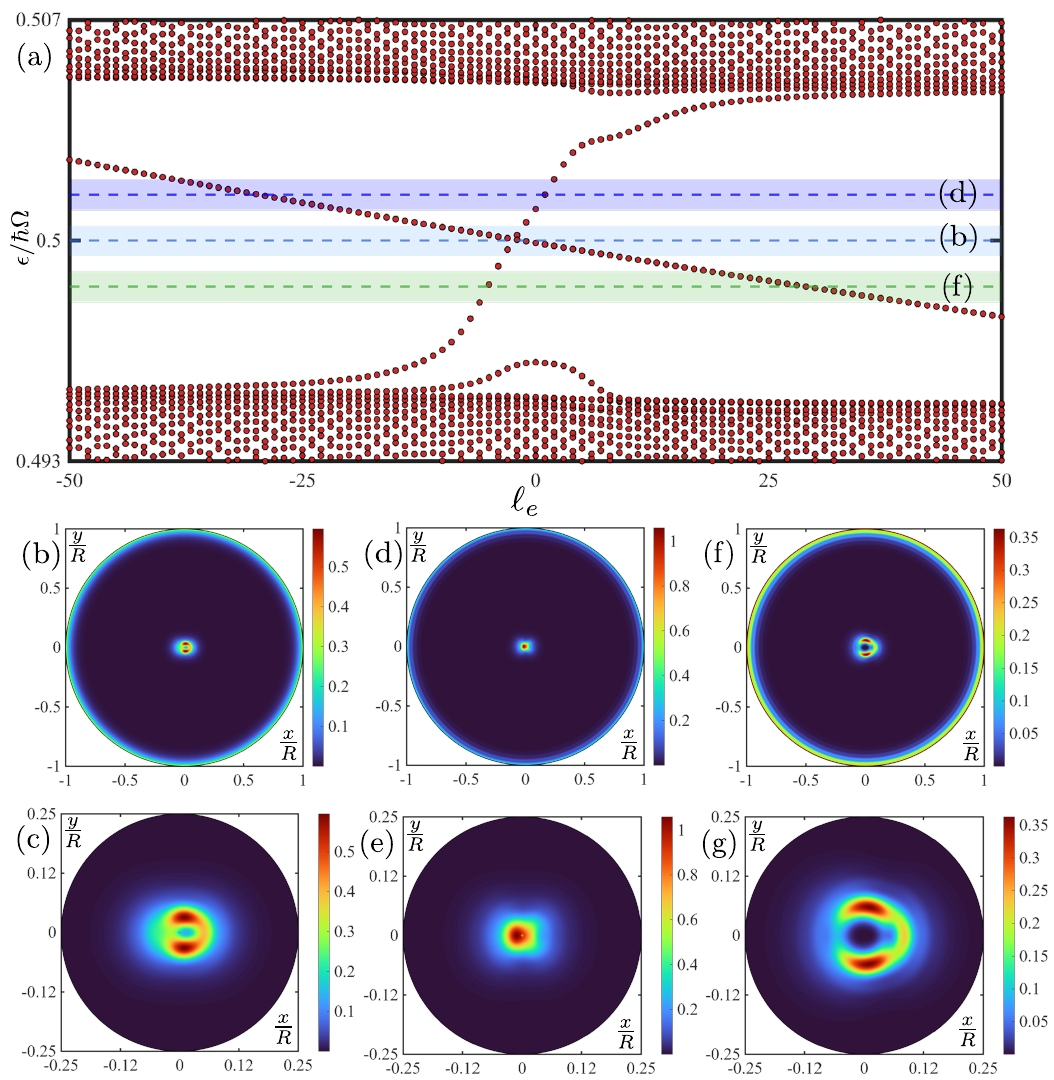}
\caption{Impurity-modified LDOS for the configuration of Fig.~\ref{Fig6}. Panel (a) shows the quasienergy spectrum for $\ell=-1$, with the probe quasienergies $E_{0}$ and broadening $\Gamma$ indicated. Panels (b), (d), and (f) show the LDOS at $E_{0}=0,\;0.003M,$ and $-0.003M$, respectively, while panels (c), (e), and (g) show the corresponding zoomed-in LDOS maps. All other parameters are as in Fig.~\ref{Fig4}.}
  \label{Fig7}
\end{figure}
For an off-center Gaussian impurity, the angular dependence of the impurity potential can be expanded in Fourier components proportional to
$e^{i\nu(\theta-\theta_0)}$, weighted by modified Bessel functions
$I_\nu(rr_0/a_0^2)$. Consequently, the impurity couples different angular-momentum sectors through angular-momentum transfer $\nu$, so that coupling between states with $j\neq j'$ requires $\nu\neq0$. As a result, the off-diagonal matrix elements $\mathcal{V}^{m'j'}_{mj}$ are controlled by the modified Bessel functions $I_{\nu\neq0}$. In particular, when the impurity is located at the vortex center, $r_0=0$, one has $I_{\nu}(0)=0$ for all $\nu\neq0$, implying $\mathcal{V}^{m'j'}_{mj}=0$ whenever $j\neq j'$. Therefore, in this rotationally symmetric limit the impurity does not mix different angular-momentum sectors, and the Floquet states remain unchanged. 

\begin{figure}[ht!] 
  \centering
  \includegraphics[width=0.5\textwidth]{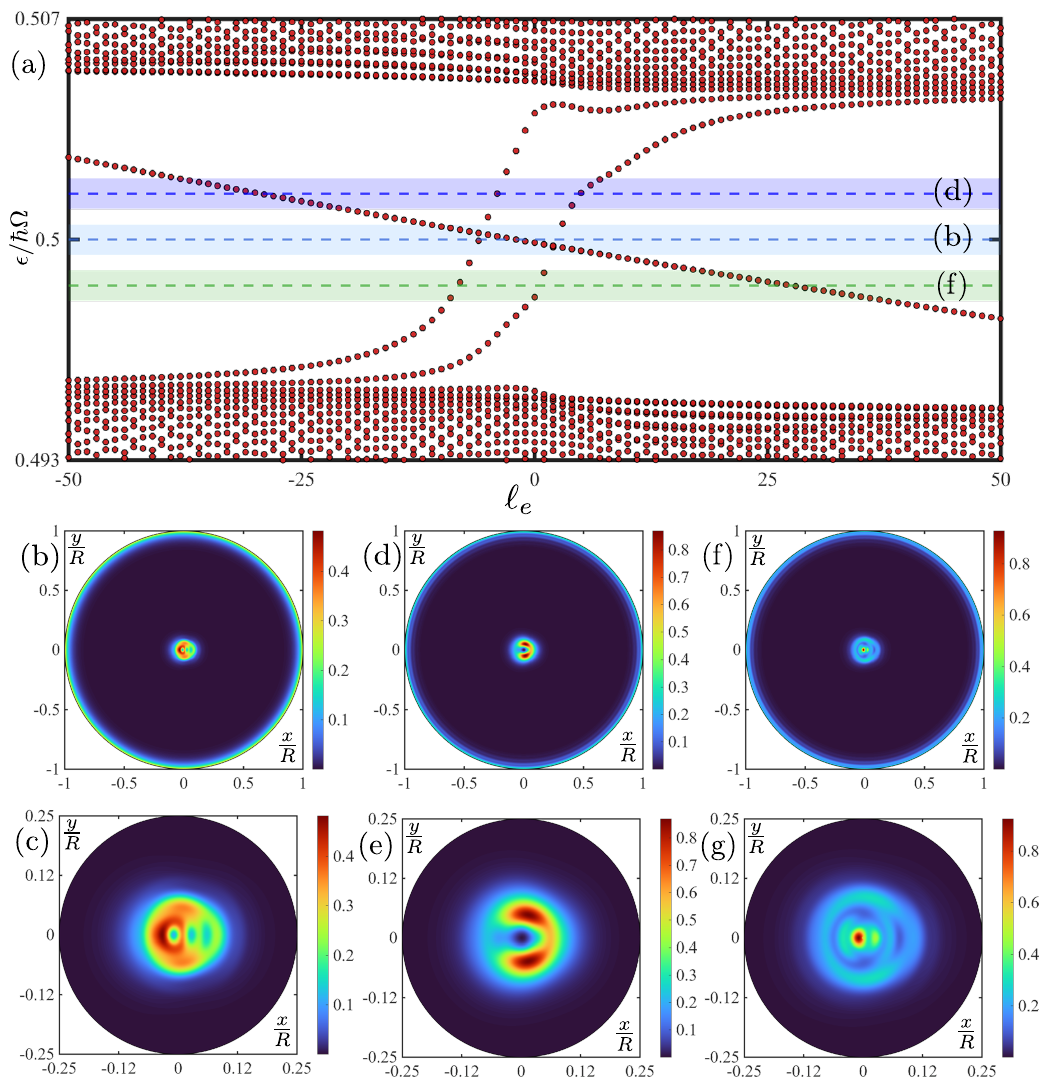}
\caption{Same as Fig.~\ref{Fig7}, but for $\ell=-2$. Panel (a) shows the impurity-modified quasienergy spectrum, while panels (b), (d), and (f) show the LDOS at $E_{0}=0,\;0.003M,$ and $-0.003M$, respectively. Panels (c), (e), and (g) show the corresponding zoomed-in LDOS maps.}
  \label{Fig7-2}
\end{figure}
In Fig.~\ref{Fig6}(a), for $\ell=-1$, the impurity-induced charge redistribution cannot be
understood as a simple rigid displacement of the vortex states away from a
repulsive potential. In the clean system, the vortex states are angular
momentum eigenstates and therefore have cylindrically symmetric ring-like
densities. The off-center Gaussian impurity breaks this rotational symmetry
and mixes nearby angular-momentum channels. The direction of the resulting
density modulation is determined by the relative weight and phase of the
admixed components. For the larger-$\ell_e$ states, such as
$\ell_e=7,8,9,\ldots$, the impurity lies on the inner side of the clean
ring-like density profile. These states are therefore mixed more strongly
with lower-$\ell_e$ states, whose radial weight overlaps more strongly with
the impurity region. The interference with these lower-$\ell_e$ components
produces a positive angular modulation on the impurity side, leading to an
apparent accumulation of charge near the impurity. This behavior is
consistent with the perturbed quasienergy, Fig.~\ref{Fig6}(b), where these states lie on the
decreasing side of the impurity-overlap peak, so that neighboring
lower-$\ell_e$ states experience larger shifts than neighboring
higher-$\ell_e$ states. By contrast, for smaller angular momenta, the balance of the admixed components is different, leading to the opposite sign of the angular modulation and hence to a
depletion of charge near the impurity. Thus the sign of the spatial
redistribution is controlled not only by the repulsive character of the
impurity potential, but also by the impurity-induced mixing between nearby
vortex states.

The hole-like states of the pristine spectrum show a complementary radial trend that can be understood from the radial organization of the in-gap spectrum in Fig.~\ref{Fig3}(a), where the vortex states form two radial ``cones'' (electron-like and hole-like). The dipole-like angular modulation remains a generic consequence of the off-center impurity, while its detailed orientation depends on the dominant hybridization partners and therefore on whether the state originates from the electron-like or hole-like branch of the pristine spectrum.

For $\ell=-2$, the impurity modifies the states in both vortex branches, as shown in Figs.~\ref{Fig6}(c) and (d). 
In this case, the real-space response is more complex. This follows from the larger number of nearby states available for impurity-induced hybridization. In first-order perturbation theory, Eq.~\eqref{correctedstate}, the corrected state is controlled by matrix elements between states with both small quasienergy separation and appreciable spatial overlap. Thus, for $\ell=-2$, the impurity can induce not only intra-branch mixing, but also inter-branch mixing, as well as vortex--bulk mixing. 
The resulting interference produces the radial oscillations and reverberation patterns visible in Figs.~\ref{Fig6}(c) and (d). In addition, states near the bulk manifold acquire extended components through vortex--bulk mixing, further enhancing these oscillatory features. 

This phenomena can be also seen in the impurity-modified LDOS, obtained by replacing the pristine quasienergies and Floquet states in Eq.~\eqref{LSOS-clean} with the impurity-corrected quasienergies, $E^{m}_{j}+\Delta E^{m}_{j}$, and the corrected states in Eqs.~\eqref{epsilon_c} and \eqref{correctedstate}.
In Figs.~\ref{Fig7}(b)--(f) [Figs.~\ref{Fig7-2}(b)--(f)], we show the LDOS for $\ell=-1$ ($\ell=-2$) the three probe quasienergies $E_{0}=0,\pm0.003M$ indicated in Fig.~\ref{Fig7}(a)[Fig.~\ref{Fig7-2}(a)]. In both cases, the LDOS near VLB center acquire intricate patterns
as shown in the zoomed-in panels (c), (e), and (g) in Fig.~\ref{Fig7} [Fig.~\ref{Fig7-2}] due to the multiple interference channels.

We also note that the topological edge states remain strongly localized at the sample boundary and are only weakly modified by the impurity, panels (b), (d), and (f) in Figs.~\ref{Fig7} and~\ref{Fig7-2}, due to their small spatial overlap with the impurity region. 


\subsection{Effects of Deviations from Circular Polarization on Quasienergy Spectrum} 
Beyond material imperfections such as impurities, deviations from CP also break total angular-momentum conservation (Sec.~\ref{FLsymmetry}). We model this effect perturbatively by taking the unperturbed drive to be circularly polarized with handedness $\tau=+1$ and adding a small component with the opposite handedness, while keeping the total light intensity fixed:
\begin{equation}\label{CPpert}
{\bm{\mathcal{{A}}}}({\bm r, t})=\Re\!\big\{{\bar{A}(r)}\,e^{i(\Omega t-\ell\phi)}
\left(\boldsymbol{\varepsilon}_\frac{\pi}{2}+\lambda\boldsymbol{\varepsilon}_{-\frac{\pi}{2}}\right)\big\},
\end{equation}
where $\boldsymbol{\varepsilon}$ is defined in Eq.~\eqref{Alight}, 
${\bar{A}(r)}=\sqrt{2}A(r)/\sqrt{(1+\lambda)^2+(1-\lambda)^2}$, and $\lambda$ parametrizes the deviation from CP, as illustrated in Fig.~\ref{Fig1}(e). The normalization of $\bar A(r)$ keeps the total light intensity fixed as the polarization is varied: $\lambda=0$ corresponds to CP, while increasing $\lambda$ continuously detunes the drive toward linear polarization, reached at $\lambda=1$. In the projected Floquet description, the opposite-handed component couples to the near-resonant subspace only through an off-resonant process, so its effective matrix element is suppressed by the small factor $\widetilde m/(2\hbar\Omega)$. We absorb this factor into $\lambda$ and therefore treat $\lambda$ as the scaled strength of the CP-detuning perturbation in the effective Hamiltonian. This allows us to capture the leading low-energy effect of polarization detuning perturbatively within the same projected Floquet model, without increasing the drive intensity or introducing additional Floquet replicas.
\begin{figure}[ht!] 
  \centering
  \includegraphics[width=0.5\textwidth]{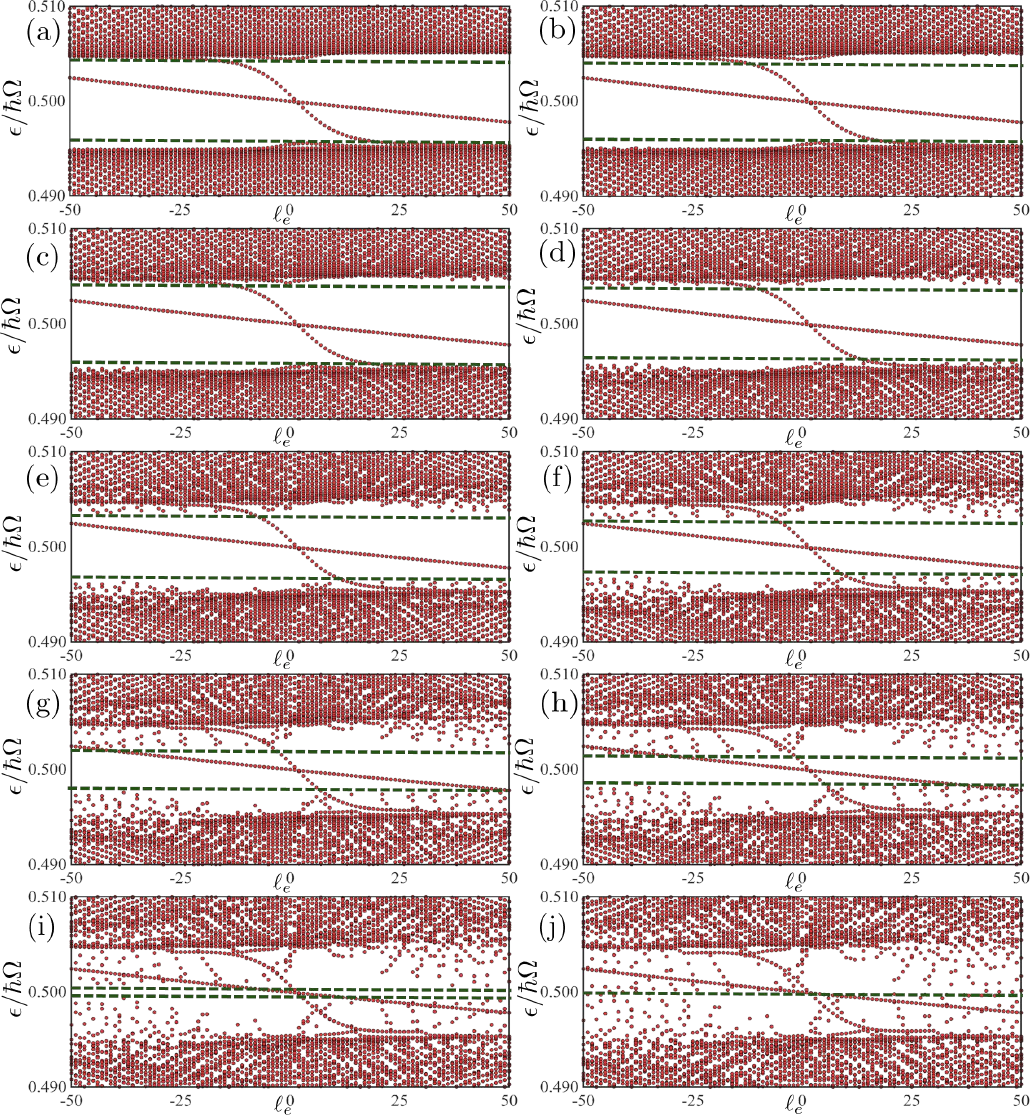}
\caption{Quasienergy spectrum of the VLB-driven system under CP detuning. Panels (a)--(j) correspond to $\lambda=0.01,0.02,\ldots,0.10$, respectively. Dashed horizontal lines mark the effective dynamical-gap edges, whose separation decreases with increasing $\lambda$. As the detuning grows, bulk-derived states enter the gap and hybridize with the vortex and edge branches. All other parameters are as in Fig.~\ref{Fig2}.}
  \label{Fig8}
\end{figure}

In the effective Floquet description, this deviation introduces a polarization-detuning perturbation,
\begin{equation}\label{HFCPpert}
\tilde{H}(\bm r)=H_{\rm F}(\bm r)+\lambda H_{\rm det}(\bm r),
\end{equation}
with
$
H_{\rm det}(\bm r)=\hat{A}(r)\sum_{\gamma=\pm}\alpha_{\gamma}\sigma_{\gamma}e^{-i\gamma\ell\theta}
$,
where $H_{\rm F}(\bm r)$ is obtained from Eq.~\eqref{effectiveH} by replacing $A(r)\rightarrow\bar{A}(r)$, and $\hat{A}(r)=ev_{\rm F}\bar{A}(r)$. Following the steady-state and perturbative framework introduced in Secs.~\ref{LDOS} and \ref{ImpurityPert}, we treat $H_{\rm det}(\bm r)$ as a perturbation to the pristine Floquet states and quasienergies in Eq.~\eqref{effective_states}. The first-order quasienergy correction vanishes after angular integration,
\begin{equation}\label{1stpert}
 \Delta E^{m,(1)}_{j}
 =\lambda\int d^2r\;\hat{\Phi}^{\rm \dag}_{m,j}(\bm r)H_{\rm det}(\bm r)\hat{\Phi}_{m,j}(\bm r)=0,
\end{equation}
because the detuning term carries angular momentum relative to the $\tau=+1$ circularly polarized basis and therefore has no diagonal matrix element within the same $j$ sector.

The leading quasienergy correction therefore appears at second order in $\lambda$. The relevant off-diagonal matrix elements are
\begin{align}\label{LambdaCP}
\Lambda^{m'j'}_{mj}
=\int d^2r\;\hat{\Phi}^{\dagger}_{m',j'}(\bm r)\,H_{\rm det}(\bm r)\,\hat{\Phi}_{m,j}(\bm r).
\end{align}
The angular integration imposes a selection rule: the opposite-spin and Floquet-harmonic component changes the total angular momentum by two units, so only matrix elements connecting $j$ to $j\pm2$ survive,
$
\Lambda^{m'j'}_{mj}\neq0\Longrightarrow
j'=j\pm2.
$
Equivalently, the nonzero matrix elements can be written as
\begin{align}\label{LambdaCP_radial}
\Lambda^{m'j'}_{mj}
=&\;\int
[\phi^{1}_{m',\uparrow,j'}(r)]^*\,\hat{A}(r)\,
\phi^{0}_{m,\downarrow,j}(r)\delta_{j',j-2}\,r\,dr \nonumber\\
&+\int
[\phi^{0}_{m',\downarrow,j'}(r)]^*\,\hat{A}(r)\,
\phi^{1}_{m,\uparrow,j}(r)\delta_{j',j+2}\,r\,dr .
\end{align}
Thus, the second-order correction to the quasienergy is
\begin{equation}\label{2ndqe}
  \Delta E^{m,(2)}_{j}
  =\lambda^2\sum_{m'}\left[
  \frac{|\Lambda^{m',j-2}_{m,j}|^2}{E^{m}_{j}-E^{m'}_{j-2}}
  +
  \frac{|\Lambda^{m',j+2}_{m,j}|^2}{E^{m}_{j}-E^{m'}_{j+2}}
  \right].
\end{equation}
Fig.~\ref{Fig8} shows the evolution of the quasienergy spectrum as the drive is detuned away from CP. Panels (a)--(j) correspond to $\lambda=0.01,0.02,\dots,0.10$, respectively. The dashed horizontal lines indicate the effective upper and lower edges of the remaining dynamical gap for each value of $\lambda$; their gradual approach toward one another provides a visual measure of the gap closing. For the smallest detunings [Figs.~\ref{Fig8}(a) and (b)], the spectrum remains close to the circularly polarized case: the vortex- and edge-state branches are still well defined inside the dynamical gap, while the bulk states remain mostly outside the gap edges. As $\lambda$ increases [Figs.~\ref{Fig8}(c)--(f)], bulk states near the upper and lower gap edges begin to move into the gap, reducing the spectral separation between the in-gap branches and the bulk continuum. For larger detuning [Figs.~\ref{Fig8}(g)--(j)], the effective gap becomes increasingly populated by bulk-derived states, and the originally isolated vortex- and edge-state branches become strongly mixed with the surrounding spectrum.

This behavior follows from the perturbative structure of the polarization-detuning term. Since the diagonal first-order correction vanishes, the leading quasienergy correction is second order in $\lambda$ and is controlled by the off-diagonal matrix elements in Eq.~\eqref{LambdaCP}. The selection rule $j\rightarrow j\pm2$ allows states with different total angular momenta to hybridize, with the largest corrections occurring when the coupled states are close in quasienergy and have appreciable spatial overlap. Therefore, vortex states lying closer to the bulk continua are affected more strongly, because their smaller quasienergy detuning from nearby bulk states enhances level repulsion. States coupled to nearby higher-energy states are pushed downward, while states coupled to nearby lower-energy states are pushed upward. This also explains the inward bending of the upper and lower ends of the vortex branches at larger $\lambda$: the upper portions are repelled downward by nearby bulk states, while the lower portions are repelled upward. As $\lambda$ grows, this level repulsion drives bulk states from both gap edges into the dynamical gap, so the gap closes by being progressively filled with bulk-derived states.

Importantly, this gap closing remains consistent with particle-hole symmetry. The polarization-detuning perturbation breaks total angular-momentum conservation, but it does not break the effective particle-hole symmetry of the Floquet Hamiltonian. Consequently, bulk states enter the gap in particle-hole-related pairs, and the spectrum remains symmetric about the Floquet-zone edge even as the dynamical gap is filled and the vortex branches lose their isolation.

\section{Discussion and Conclusion}\label{Disc}

We studied impurity scattering and polarization detuning in finite-size vortex-light-beam--driven massive Dirac systems. As shown in Ref.~\cite{Lauren2025}, for circularly polarized VLBs, the interplay between the light's spin and orbital angular momenta generates a dynamical gap near the one-photon resonance, where topological edge states coexist with photoinduced multiply quantized vortex states. The VLB OAM fixes the number and directionality of the vortex branches, while the finite geometry allows edge, bulk, and vortex states to be resolved within the same quasienergy spectrum.

The quasienergy spectrum and states are constrained by symmetry. In the pristine CP case, Floquet total angular momentum labels the states, while effective particle-hole symmetry fixes their pairing in quasienergy and angular momentum. This accounts for the even--odd distinction in $|\ell|$, the particle-hole symmetry center $\ell_c$, and the identical real-space densities of particle-hole-related vortex states. These symmetry constraints also clarify the perturbative response: off-center scalar impurities break rotational symmetry and particle-hole pairing through angular-momentum mixing, whereas polarization detuning breaks angular-momentum conservation but preserves particle-hole symmetry, leading to a symmetric filling of the dynamical gap.

Together, impurities and polarization detuning show that the VLB-induced vortex and edge structure is robust but sensitive to angular-momentum mixing. Off-center impurities reshape vortex states and LDOS through spatial overlap with localized states, while deviations from CP mix sectors through $j\rightarrow j\pm2$, producing second-order level repulsion that is strongest near the bulk continua. These effects modify the spectra and real-space profiles without obscuring the organizing role of vorticity and, when preserved, particle-hole symmetry. 


These results are directly relevant for experiments on realistic driven materials, particularly thin films of topological insulators, where massive Dirac surface states and finite-size geometries provide a natural platform~\cite{thin2,thin4,thin3}. The persistence of the vortex and edge signatures in the presence of localized impurities or extended impurity clusters, as well as under polarization deviations as large as $10\%$ from CP, suggests that VLB-induced states should remain experimentally resolvable. Spatial probes such as STM~\cite{TimeSTM}, and especially time-resolved lightwave-driven scanning tunneling spectroscopy, could detect the predicted LDOS signatures of both the topological edge states and the multiply quantized light-induced vortex states, even in samples with impurity imperfections and small deviations from ideal circular polarization.

\begin{acknowledgements}
This work was supported by the U.S. Department of Energy, Office of Science, Basic Energy Sciences, under Award \# DE-SC0025703.
\end{acknowledgements}    

\bibliography{References.bib}

\end{document}